\pdfoutput=1
\documentclass{article}
\usepackage[utf8]{inputenc}
\usepackage{dcolumn,booktabs}
\newcolumntype{d}[1]{D..{#1}}
\usepackage[section]{placeins}
\usepackage{lscape} 
\usepackage{amsmath}
\usepackage{fullpage}
\usepackage{microtype}
\usepackage[titletoc,toc,page]{appendix}

\usepackage[table, svgnames]{xcolor}
 
\usepackage{array, makecell}
\usepackage{multirow}
\usepackage{authblk}
\usepackage{subfig}
\usepackage{tcolorbox}

\title{\textbf{Clinical prediction system of complications among COVID-19 patients: a development and validation retrospective multicentre study}}

\author[1]{Ghadeer O. Ghosheh}
\author[1]{Bana Alamad}
\author[1]{Kai-Wen Yang}
\author[2]{Faisil Syed}
\author[1]{Nasir Hayat}
\author[2]{Imran Iqbal}
\author[2]{Fatima Al Kindi}
\author[2]{Sara Al Junaibi}
\author[2]{Maha Al Safi}
\author[1]{Raghib Ali}
\author[3]{Walid Zaher}
\author[2*]{Mariam Al Harbi}
\author[1*$\dagger$]{Farah E. Shamout}
\affil[1]{Engineering Division, NYU Abu Dhabi}
\affil[2]{Abu Dhabi Health Services}
\affil[3]{G42 Healthcare}
\affil[*]{Joint supervision}
\affil[$\dagger$]{fs999@nyu.edu}
\date{}
\usepackage{graphicx}

\begin{document}

\maketitle
\vspace{-15mm}
%TRIPOD: Provide a summary of objectives, study design, setting, participants, sample size, predictors, outcome, statistical analysis, results, and conclusions
\section*{Abstract} 

\subsubsection*{Background}
Existing prognostic tools mainly focus on predicting the risk of mortality among patients with coronavirus disease 2019 (COVID-19). However, clinical evidence suggests that COVID-19 can result in non-mortal complications that affect patient prognosis. To support patient risk stratification, we aimed to develop a prognostic system that predicts complications common to COVID-19. 

\subsubsection*{Methods}
In this retrospective study, we used data collected from 3,352 COVID-19 patient encounters admitted to 18 facilities between April 1 and April 30, 2020, in Abu Dhabi (AD), United Arab Emirates. The hospitals were split based on geographical proximity to assess for our proposed system's learning generalizability, AD Middle region and AD Western \& Eastern regions, A and B, respectively. Using clinical data collected during the first 24 hours of admission, the machine learning-based prognostic system predicts the risk of developing any of seven complications during the hospital stay. The complications include secondary bacterial infection, Acute Kidney Injury (AKI), Acute Respiratory Distress Syndrome (ARDS), and elevated biomarkers linked to increased patient severity, including d-dimer, interleukin-6, aminotransferases, and troponin. During training, the system applies an exclusion criteria, hyperparameter tuning, and model selection for each complication-specific model. We assessed its performance using the area under the receiver operating characteristic curve (AUROC) and the area under the precision recall curve.

\subsubsection*{Findings}
The system achieves good accuracy across all complications and both regions. In test set A (587 patient encounters), the system achieves 0.91 AUROC for AKI and $>0.80$ AUROC for most of the other complications. In test set B (225 patient encounters), the respective system achieves $\geq0.90$ AUROC for AKI, elevated troponin, and elevated interleukin-6, and $>0.80$ AUROC for most of the other complications. The best performing models, as selected by our system, were mainly gradient boosting models and logistic regression. 

\subsubsection*{Interpretation}
Our results show that a data-driven approach using machine learning can predict the risk of such complications with high accuracy. Such an early risk prediction system could have real-life impact in supporting clinical decision-making for COVID-19 patients.
%We present novel risk prediction benchmarks for COVID-19 patients and open-access pipeline can also be easily adapted to other patient cohorts.

\subsubsection*{Funding}
New York University Abu Dhabi.

% Contributions: 
% (1) there's benefit in using ML to predict such complications, which were not commonly investigated previously. ML is promising here
% (2) establish benchmark risk prediction scores for COVID-19 patients
% (3) novel framework can be easily reused for other cohorts

\vspace{1 cm}
    \begin{sloppypar}
\begin{tcolorbox}
[%
%   experimental split,
%   bicolor,
  collower=white,
%   halign=center,
  valign lower=top,
  toptitle=1.5mm,
  title={Research In Context},
%   colbacklower=black!70,
  ]%
    % \vspace{-0.45cm}
    % \begin{tabular}{|p{\textwidth}|}
    % \midrule
        \textbf{Evidence before this study.}
        
        We conducted an extensive literature review between May 30, 2020, and July 10, 2020, to assess published work until end of July, 2020, using the key terms “coronavirus”, “COVID-19”, “prognosis”, “prediction model”, and “machine learning”. Peer reviewed papers obtained from PubMed and EMBASE in addition to preprints obtained from MedRxiv and BioRxiv were included in our review. 
        We excluded studies that did not incorporate any machine learning, studies that focused on diagnosis rather than prognosis, and studies that relied on imaging data. The majority of published work for COVID-19 patients propose to predict admission to the intensive care unit, mortality, prolonged hospital stay or the need for mechanical ventilation. There was a clear gap in the machine learning literature for the simultaneous prediction of non-mortal complications among COVID-19 patients, despite the vast clinical evidence showing that COVID-19 can result in such complications. 

        \textbf{Added value of this study.} 
        
        We analyzed data extracted from electronic health records for a large cohort of COVID-19 patients (3,352 COVID-19 patient encounters) admitted to hospitals in Abu Dhabi, United Arab Emirates, to develop a risk prediction system for seven common complications among COVID-19 patients. We focus on the accurate prediction of seven complications since they are precursors to widely studied adverse events, including mortality.  This is the first study to study such complications and to present the patient cohort in the UAE, as most previous studies have focused on European or Chinese patient cohorts. Our approach achieves good accuracy across all complications and can be easily adapted to external patient cohorts.  

        \textbf{Implications of all available evidence.}
        
        We present data-driven system that uses machine learning to predict seven complications indicative of COVID-19 patient severity. The results highlight the promise of machine learning to support clinical decision-making and hospital resource management. 
\end{tcolorbox}

\end{sloppypar}
\newpage

\section{Introduction}

The Severe Acute Respiratory Syndrome Coronavirus 2 (SARS-CoV-2) has led to a global health emergency since the emergence of the coronavirus disease 2019 (COVID-19). Despite containment efforts, more than 55 million confirmed cases have been reported globally, of which 157,785 cases are in the United Arab Emirates (UAE) as of November 21, 2020~\cite{dong2020interactive}. Due to unexpected burdens on healthcare systems, identifying high risk groups using prognostic models has become vital to support patient triage. 

Most of the recently published prognostic models focus on predicting mortality, the need for intubation, or admission into the intensive care unit~\cite{critical}. While the prediction of such adverse events is important for patient triage, clinical evidence suggests that COVID-19 may result in a variety of complications in organ systems that may eventually lead to mortality~\cite{WHO_lancet}. For example, Acute Respiratory Distress Syndrome (ARDS) related pneumonia has been reported as a major complication of COVID-19~\cite{2020FeiTingRonghui}. Other studies reported alarming percentages among hospitalized COVID-19 patients that have developed hematological complications~\cite{ddimer2}, organ dysfunction~\cite{kindey2}, or secondary bacterial infection~\cite{2020FeiTingRonghui}. Table~\ref{tab:Lit_review} summarizes key studies that reported diagnosed complications or biomarkers which may lead to severe complications across different COVID-19 patient populations. Those findings suggest a pressing need for the development and validation of a prognostic system that predicts such complications in COVID-19 patients to support patient management.

Here, we address this need by proposing an automated prognostic system that learns to predict a variety of non-mortal complications among COVID-19 patients admitted to the Abu Dhabi Health Services (SEHA) facilities, UAE. The system uses multi-variable data collected during the first 24 hours of the patient admission,  including vital-sign measurements, laboratory-test results, and baseline information. We particularly focus on seven complications based on the clinical evidence presented in Table~\ref{tab:Lit_review}, which are either based on clinical diagnosis or biomarkers that are indicative of patient severity.  To allow for reproducibility and external validation, we made our code and a test set publicly available at: \url{ https://github.com/nyuad-cai/COVID19Complications}.

 \begin{table}[ht]
 \centering
 \small
  \caption{\small Summary of clinical studies reporting various non-mortal complications in patients with confirmed COVID-19 diagnosis. The alarming incidence rates suggest a pressing need for developing a clinical decision support system that predicts such complications.} 
        \begin{tabular}{lcclc}
            \toprule
            \textbf{Complication} & \textbf{Cohort Size} & \textbf{Incidence rate} & \textbf{Location} & \textbf{References} \\
          \midrule
             Elevated Troponin & \multicolumn{1}{l}{614} & \multicolumn{1}{l}{45.3\%} &
             \multicolumn{1}{l}{Italy} & \multicolumn{1}{c}{\cite{elevated_troponin1} }  \\
                                 & \multicolumn{1}{l}{1527} & \multicolumn{1}{l}{8.0\%}
                                 & \multicolumn{1}{l}{China}
                                 & \multicolumn{1}{c}{\cite{MI_china}}
                                  \\\midrule
                                  
            Elevated D-Dimer & \multicolumn{1}{l}{248} & \multicolumn{1}{l}{74.6\%} &
            \multicolumn{1}{l}{China} &
            \multicolumn{1}{c}{\cite{ddimer1}} \\
                                 & \multicolumn{1}{l}{2377} & \multicolumn{1}{l}{76.0\%}
                                 & \multicolumn{1}{l}{United States} & \multicolumn{1}{c}{\cite{ddimer2}} 
                                 \\
                                 \midrule
          Elevated Aminotransferases  & \multicolumn{1}{l}{105} & \multicolumn{1}{l}{21.0\%} & \multicolumn{1}{l}{China} & \multicolumn{1}{c}{\cite{amino2}}  \\
                                    &
        \multicolumn{1}{l}{5700} & \multicolumn{1}{l}{39.0\% \& 58.5\%$^*$} & \multicolumn{1}{l}{United States} & \multicolumn{1}{c}{\cite{amino1}}  \\\midrule
              Elevated Interleukin-6 
          & \multicolumn{1}{l}{728} & \multicolumn{1}{l}{16.5\%} &
          \multicolumn{1}{l}{China} &
          \multicolumn{1}{c}{\cite{elevated_il6}}
         \\
         %& \multicolumn{1}{l}{584} & \multicolumn{1}{l}{56.5\%} & \multicolumn{1}{l}{Spain} & \multicolumn{1}{c}{\cite{il6_2}}  \\
            \midrule
 Secondary Bacterial Infection & 
            \multicolumn{1}{l}{191} & \multicolumn{1}{l}{15.0\%} & \multicolumn{1}{l}{China} & \multicolumn{1}{c}{\cite{2020FeiTingRonghui}}  \\
                        &     
            \multicolumn{1}{l}{338} & \multicolumn{1}{l}{5.6\%$\dagger$} & \multicolumn{1}{l}{United States} & \multicolumn{1}{c}{\cite{2020GoyalChoi}}   \\
            \midrule
           
            Acute Kidney Injury 
            & \multicolumn{1}{l}{5449} & \multicolumn{1}{l}{36.6\%}&
            \multicolumn{1}{l}{United States}&
            \multicolumn{1}{c} {\cite{kindey2}} \\
            &
            \multicolumn{1}{l}{98} & \multicolumn{1}{l}{9.2\%} & \multicolumn{1}{l}{South Korea} & \multicolumn{1}{c}{\cite{ards1}} 
                                 \\\midrule
          \multirow{2}{*}{\begin{tabular}[c]{@{}l@{}}Acute Respiratory Distress \\Syndrome \end{tabular}} &    \multicolumn{1}{l}{191} & \multicolumn{1}{l}{31.0\%} & \multicolumn{1}{l}{China} & \multicolumn{1}{c}{\cite{2020FeiTingRonghui}}  \\
          & 
                                  \multicolumn{1}{l}{98} & \multicolumn{1}{l}{18.4\%} & \multicolumn{1}{l}{South Korea} & \multicolumn{1}{c}{\cite{ards1}} \\

    & \multicolumn{1}{l}{1099} & \multicolumn{1}{l}{3.4\%} & \multicolumn{1}{l}{China} & \multicolumn{1}{c}{\cite{guan2020clinical}} \\
            \bottomrule
         \multicolumn{5}{l}{\footnotesize $^*$ This study reported a 39.0\% incidence rate for alanine aminotransferase and \& 58.5\% for aspartate aminostransferase.} \\ 
         \multicolumn{5}{l}{\footnotesize $\dagger$ This was specifically reported for bacteremia.}
        \end{tabular}
    \label{tab:Lit_review}

\end{table}

\section{Methods}
We reported this study following the TRIPOD guidance~\cite{collins2015transparent}.
% word count goal: 550 words or less for study design and population
% word count goal for the rest of the methods section: 650 words or less

\subsection{Data source}

%TRIPOD: Describe the study design or source of data (for example, randomised trial, cohort, or registry data), separately for the development and validation data sets, if applicable
%timeline (specify key dates), no. of patients, no. of admissions, no. of hospitals and their location etc. Describe the study design and source of data (for example, randomised trial, cohort, or registry data), separately for the development and validation data sets. Mention eligibility criteria for participants Give details of treatments received (appendix ?)\\ }
%TRIPOD: 1) Explain how the study size was arrived at. 2) Describe how missing data were handled (for example, complete-case analysis, single imputation, multiple imputation) with details of any imputation method

\begin{figure}[t!]
\setlength{\tabcolsep}{-30pt} 
    \centering
    \begin{tabular}{cc}
    \includegraphics[width=0.5\linewidth]{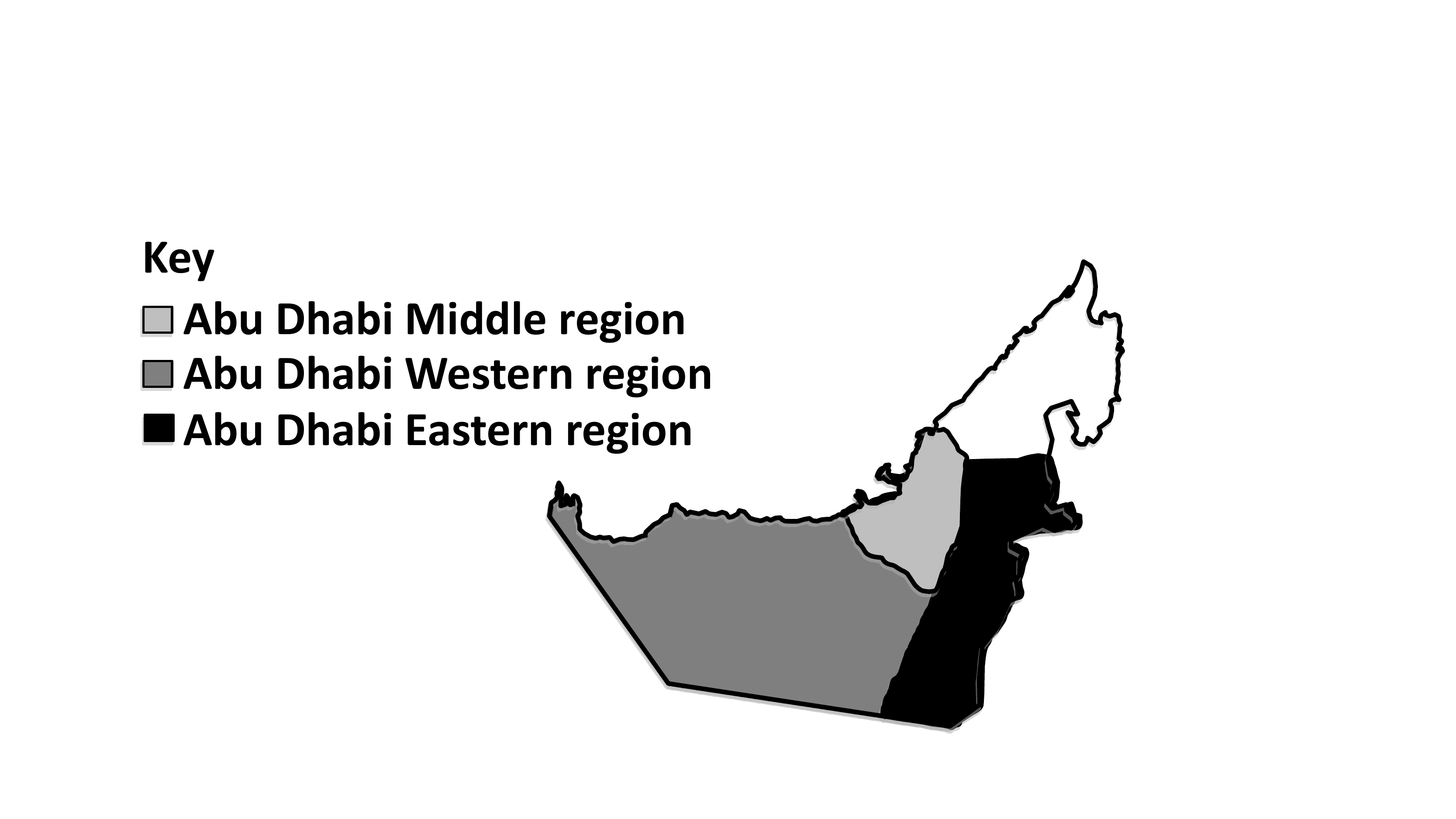} &
    \includegraphics[width=0.7\linewidth]{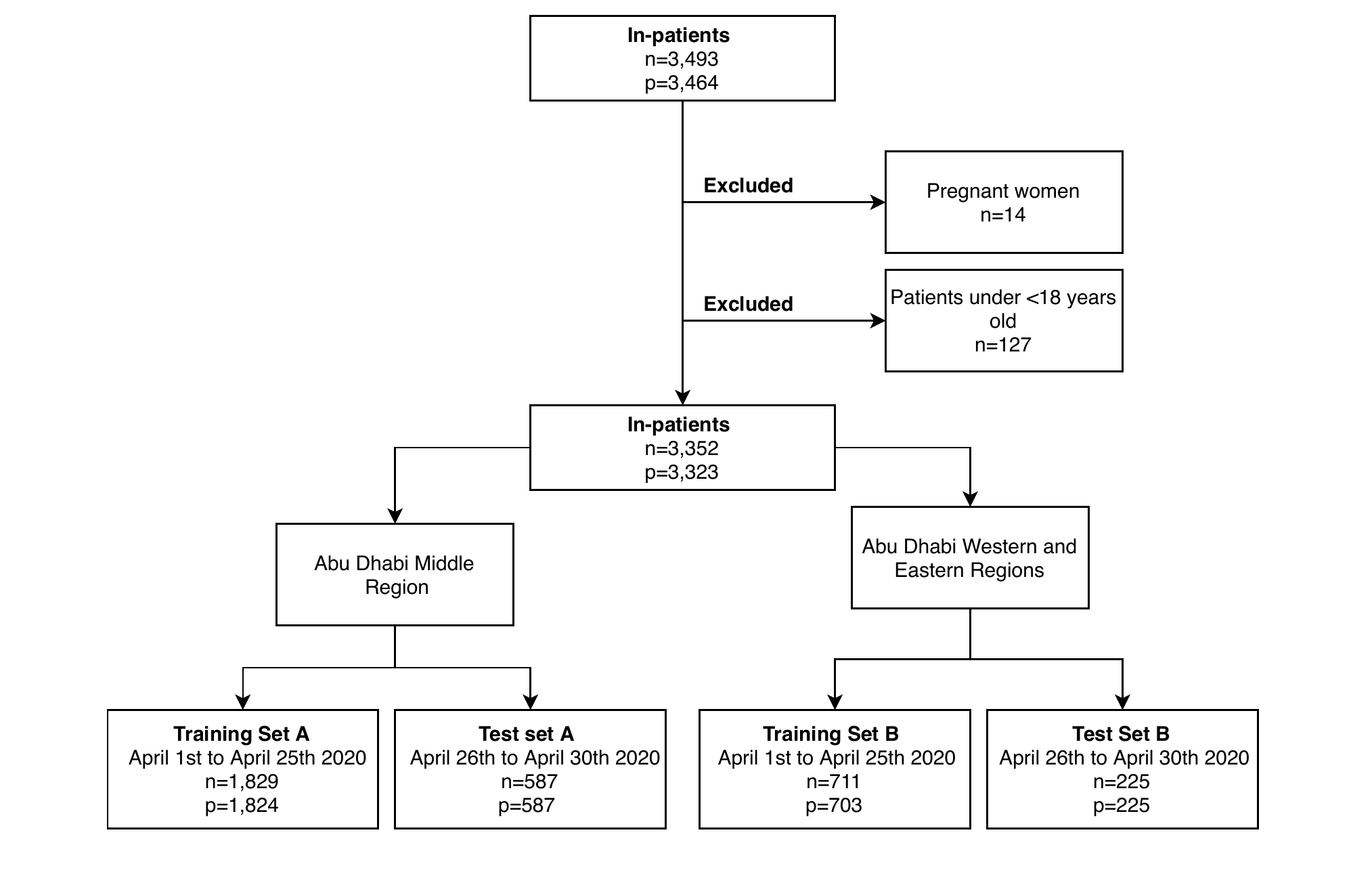} \\
    (a) & (b) \\
    \end{tabular}
    \vspace{-2mm}
    \caption{\small (a) The UAE map showcasing the location of the healthcare facilities included in this study. (b) Flowchart for the overall dataset showing how the inclusion and exclusion criteria were applied to obtain the final training and test sets, where $n$ represents the number of patient encounters, and $p$ represents the number of unique patients.}
    \label{fig:cohort}
\end{figure}

This study is a retrospective multi-centre study that includes anonymized data recorded within 3,493 COVID-19 hospital encounters at 18 Abu Dhabi Health Services (SEHA) healthcare facilities in Abu Dhabi, United Arab Emirates. The study received approval by the Institutional Review Board (IRB) from the Department of Health (Ref: DOH/CVDC/2020/1125) and New York University (Ref: HRPP-2020-70). There were 9 facilities in the Middle region, which includes the capital city, and 9 facilities in the Eastern and Western regions. Those regions are highlighted in Figure~\ref{fig:cohort}(a). Figure~\ref{fig:cohort}(b) shows the flowchart as we applied the exclusion criteria to obtain the final data splits. We excluded 127 non-adult encounters and 14 pregnant encounters and split the dataset into training and test sets. Training set A consisted of 1,829 encounters recorded in the Middle region between April 1, 2020 and April 25, 2020. To evaluate for temporal generalizability, test set A included 587 encounters recorded in the Middle region between April 26, 2020 and April 30, 2020. Training set B included 711 encounters admitted to the Eastern and Western regions between April 1, 2020 and April 25, 2020 and test set B included 225 encounters admitted to the same hospitals between April 26, 2020, and April 30, 2020.

\subsection{Defining and labeling of complications}
Based on clinical evidence and in collaboration with clinical experts, we focused on predicting seven complications, including three clinically diagnosed events such as secondary bacterial infection (SBI), Acute Kindey Injury (AKI) ~\cite{khwaja2012kdigo} and ARDS ~\cite{2012ARDS_Berlin} and four biomarkers that may be indicative of patient severity. In particular, among COVID-19 patients, elevated troponin reflects myocardial injury and has been reported to be associated with a higher risk of mortality~\cite{elevated_troponin1}, elevated d-dimer is associated with thrombotic events~\cite{ddimer1}, elevated interleukin-6 is a proinflammatory cytokine that has been shown to be associated with disease severity and in-hospital mortality~\cite{elevated_il6}, and elevated aminotransferases have been reported to be associated with liver Injury ~\cite{amino1}. For each patient encounter in the training and test sets, we identified the first occurrence (i.e., date and time), if any, of each complication based on the criteria shown in Table~\ref{tab:definitions}. The biomarkers-based complications are defined based on elevated laboratory-test results, SBI is defined based on positive cultures, AKI is defined based on the KDIGO classification criteria~\cite{khwaja2012kdigo}, and ARDS is defined based on the Berlin definition~\cite{2012ARDS_Berlin}, which required the processing of free-text chest radiology reports. Further details on the processing of those reports is described in Supplementary Section~\ref{apd:groundtruth-labels}.

\begin{table}[t!]
\caption{\small Criteria used to define the occurrence of the complications that our system aims to predict.}
\label{atts}
\centering
 \resizebox{1.0\linewidth}{!}{
\begin{tabular}{l p{12.0cm} c}
%\hline
\toprule
\textbf{Complication} & \textbf{Definition} & \textbf{Reference} \\
\midrule
% \textbf{Hematological} & &\\
Elevated Troponin & Troponin T $\geq$ 14 ng/L & ~\cite{troponins} \\
\midrule
  Elevated D-Dimer  &  D-Dimer $\geq$ 500 ng/mL & ~\cite{features_PCT} \\
 \midrule
 % \textbf{Organ dysfunction} &&\\
Elevated Aminotransferases & AST $\geq$ 40 U/l \textbf{AND} ALT $\geq$ 40 U/l &  * \\
\midrule
Elevated Interleukin-6 & Interleukin-6 $\geq$ 8.43 pg/mL & *\\
 \midrule
    SBI & Positive blood, urine, throat or sputum cultures within 24 hours of sample collection & *  \\

\midrule
 AKI &  Based on the Kidney Disease Improving Global Guidelines (KDIGO) classification, increase in Serum Creatinine by $\geq$ 0.3mg/dl within 48 hours 
 \newline
 \textbf{OR}
  \newline
 increase in Serum Creatinine by $\geq$ to 1.5 times
\newline
 \textbf{OR}
  \newline
  Urine volume $<$ 0.5ml/kg/hr for 6 hours
 & ~\cite{khwaja2012kdigo}
\\
\midrule
\multirow{2}{*}{\begin{tabular}[c]{@{}l@{}}ARDS \end{tabular}} & Based on the Berlin definition, presence of bilateral opacity in radiology reports\newline \textbf{AND} \newline Oxygenation: PaO$_2$/FiO$_2$ $\leq$300 mm Hg \newline \textbf{AND} \newline Timing: $\leq$ one week \newline \textbf{AND} \newline Origin: pulmonary &  ~\cite{2012ARDS_Berlin}\\
   
\bottomrule
\multicolumn{3}{l}{\footnotesize * Based on SEHA's clinical standards.}

\end{tabular}
}
 \label{tab:definitions} 
\end{table}

\subsection{Input features}
We considered data recorded within the first 24 hours of admission as input features to our predictive models. This data included continuous and categorical features related to the patient baseline information and demographics, vital signs, and laboratory-test results. Within the patient's baseline and demographic information, age and body mass index (BMI) were treated as continuous features, while sex, pre-existing medical conditions (i.e., hypertension, diabetes, chronic kidney disease, and cancer), and symptoms recorded at admission (i.e., cough, fever, shortness of breath, sore throat, and rash) were treated as binary features. 

%Preliminary data preprocessing was first applied to the categorical features to remove spaces and capitalization inconsistencies and then they were encoded using label encoding.

As for the vital-sign measurements and laboratory-test results, we excluded any variable that was used to define the presence of any complication in order to avoid label leakage. In particular, we considered seven continuous vital-sign features, including systolic blood pressure, diastolic blood pressure, respiratory rate, peripheral pulse rate, oxygen saturation, auxiliary temperature, and the Glasgow Coma Score, and 19 laboratory-test results, including albumin, activated partial thromboplastin time (APTT), bilirubin, calcium, chloride, c-reactive protein, ferritin, hematocrit, hemoglobin, international normalized ratio (INR), lactate dehydrogenase (LDH), lymphocytes count, prothrombin time, procalcitonin, sodium, red blood cell count (RBC), urea, uric acid, and neutrophils count. All vital-sign measurements and laboratory-test results were processed into minimum, maximum, and mean statistics. We also defined seven binary input features to represent whether a complication had occurred within the first 24 hours of admission, to allow the models to learn from any dependencies between the complications. 
% We summarized patient demographics, prevalence of the complications, and the distributions of the input features across the training and test sets. 

\subsection{Predictive modeling}
The proposed system predicts the risk of developing each of the complications during the patient's stay after 24 hours of admission. This is represented by a vector \textbf{y} consisting of 7 risk scores, where each risk score is computed by a complication-specific model, such that 
\[
   \textbf{y}= \begin{bmatrix}
      
         y\textsuperscript{El. troponin},
         y\textsuperscript{El. d-dimer},
         y\textsuperscript{El. aminotransferases}, 
         y\textsuperscript{El. interleukin-6}, 
         y\textsuperscript{SBI},
         y\textsuperscript{AKI},
         y\textsuperscript{ARDS}
         \end{bmatrix},
\]
where $y^{complication} \in [0,1]$. 

\begin{figure}[ht!]
    \centering
    \includegraphics[width=0.9\linewidth]{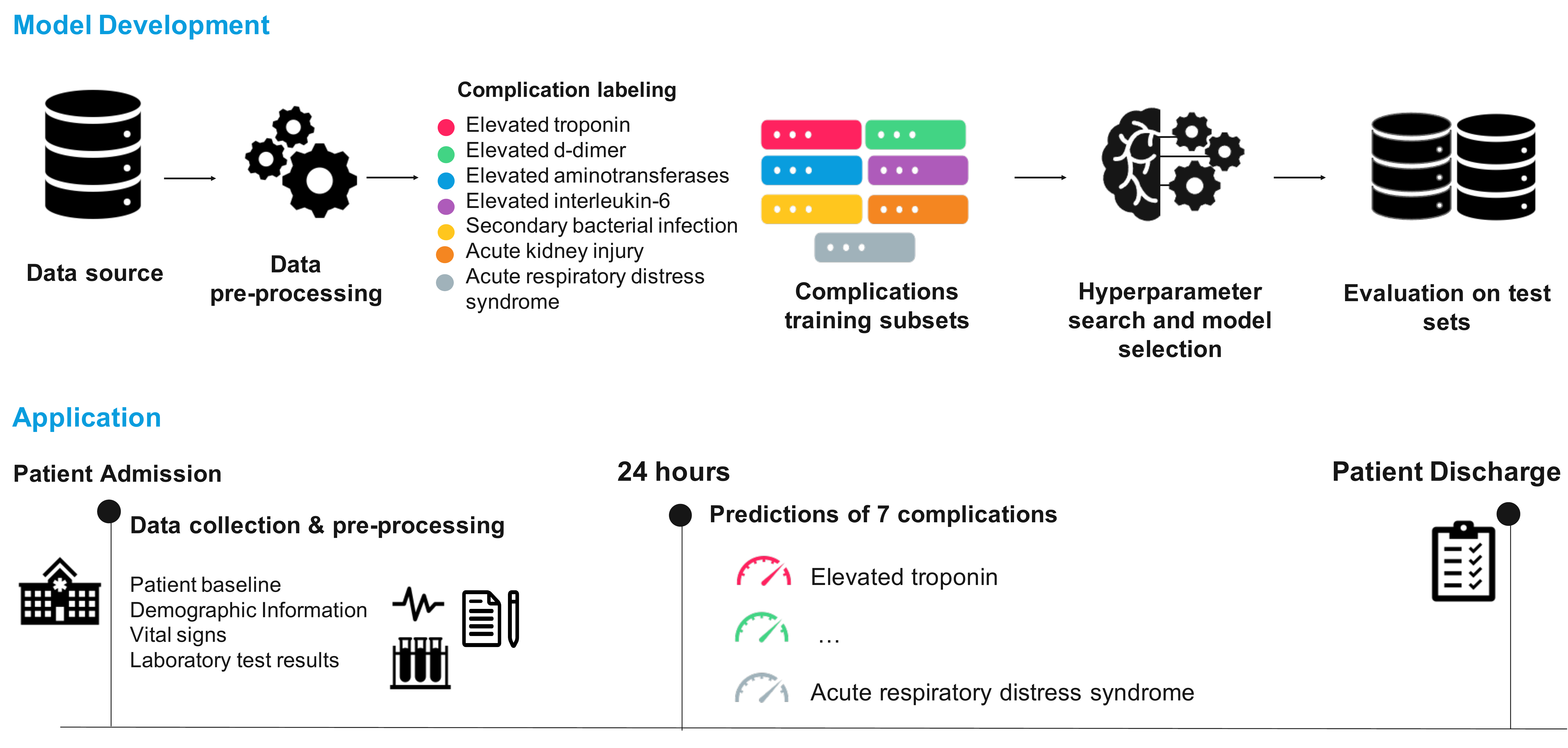}
    \caption{\small Overview of our proposed model development approach and expected application in practice. In the first row, we developed our complication-specific models by first preprocessing the data, identifying the occurrences of the complications based on the criteria shown in Table~\ref{tab:definitions}, training and selecting the best-performing models on the validation set, and then evaluating the performance on the test set, retrospectively. As for deployment, we expect our system to predict the risk of developing any of the seven complications for any patient after 24 hours of admission. }
    \label{fig:overview}
\end{figure}

The overall workflow of the model development is depicted in Figure~\ref{fig:overview}. For each complication-specific model, we excluded from its training and test sets patients who developed that complication prior to the time of prediction. For AKI, we also excluded patients with chronic kidney disease. Then for each complication, our system trains five model ensembles based on five types of base learners: logistic regression (LR), k-nearest neighbors (KNN), support vector machine (SVM), multi-layer perceptron (MLP) and a light gradient boosting model (LGBM). Missing data was imputed using median imputation for all models except for LGBM, which can natively learn from missing data, and the data was further scaled using min-max scaling for LR and MLP and standard scaling for SVM and KNN.

For each type of base learner, the system performs a stratified k-folds cross-validation using the complication's respective training set with $k=3$. We performed random hyperparameter search for each base learner's hyperparameters~\cite{bergstra2012random} over 20 iterations, resulting with 3 trained models per hyperparameter set. The hyperparameter search ranges are described in Supplementary Section~\ref{apd:parameters}. We selected the top two sets of hyperparameters that achieved the highest average area under the receiving operator characteristic curve (AUROC) on the validation sets, resulting with 6 trained models per ensemble. Then, we selected the ensemble that achieves the highest average AUROC on the validation set. Each model within the selected ensemble was further calibrated using isotonic regression on its respective validation set to ensure non-harmful decision making \cite{van2016calibration}, except for the LR models. The final prediction of each complication consisted of an average of the predictions of all calibrated base learners per ensemble.  

To understand what input features were most predictive of each complication, we performed post-hoc feature importance analysis using the tree SHapley Additive exPlanations (SHAP)~\cite{lundberg2020local}. All analysis was performed using Python (version 3.7.3), LR, KNN, SVM, and MLP models were trained using the Python \texttt{scikit-learn} package and the LGBM models were trained using the \texttt{LightGBM} package~\cite{LGBM}. 
 
% \subsection{Statistical Analysis}
% AUROC/AUPRC
% confidence intervals
%type of model? focus on interpretability Specify type of model, all model-building procedures (including any predictor selection), and method for internal validation For validation, describe how the predictions were calculated Specify all measures used to assess model performance and, if relevant, to compare multiple models Describe any model updating (for example, recalibration) arising from the validation, if done For validation, identify any differences from the development data in setting, eligibility criteria, outcome, and predictors\\
%\subsubsection{Ensembled severity scores}
\subsection{Statistical Analysis}
We evaluated each complication ensemble using the AUROC and the area under the precision recall curve (AUPRC) on the test set. Confidence intervals for all of the evaluation metrics were computed using bootstrapping with 1,000 iterations \cite{diciccio1996bootstrap}. We also assessed the calibration of the ensemble, after post-hoc calibration of its trained models, using reliability plots and reported calibration intercepts and slopes \cite{van2016calibration}.

\subsection{Role of Funding Source}
The funding source had no role in the study design or data analysis. The study was performed by all co-authors who had access to the anonymized dataset.

\section{Results}

\begin{table}[t!]                         
\centering
    \caption{\small Summary of the baseline characteristics of the patient cohort in the training sets and test sets and the prevalence of the predicted complications. Note that n represents the total number of patients while \% is the proportion of patients within the respective dataset.} 
    \resizebox{1.0\linewidth}{!}{
    \begin{tabular}{lllll}
    \toprule
     & \textbf{Training set A} & \textbf{Test set A}  & \textbf{Training set B} & \textbf{Test set B}\\

    \midrule
      \textbf{Patient Cohort} \\
    %\toprule
    % General Demographic & & &\\
     \midrule

     Encounters, n & 1829 & 587 & 711 & 225\\
      Age, mean (IQR) & 41.7 (17.0) & 45.5 (18.0) & 39.3 (17.0) & 42.7 (20.0)\\ 
     Male, n (\%) & 1582 (86.5) & 522 (88.9) & 622 (87.5) & 191 (84.8)\\
     %Female, n (\%)  & 247 (13.5) & 65 (11.1) & 89 (12.5)  & 34 (15.1)\\ 
    %  Ethnicity &&&&\\
     Arab, n (\%) & 295 (16.1)   & 89 (15.2)  & 120 (16.9) & 43 (19.1)\\
    % ~~~~~Emirati (n, \%) & 89 (4.88\%) & 26 (4.41\%) & 24 (2.58\%) &\\
    % ~~~~~Egyptian (n, \%) & 71 (3.89\%) & 32 (5.42\%) & 39 (4.18\%) &\\
    % ~~~~~Jordanian (n, \%) & 25 (1.37\%) & 7 (1.19\%) & 12 (1.29\%) &\\
    % ~~~~~Sudanese (n, \%) & 24 (1.32\%) & 5 (0.85\%) & 18 (1.93\%) &\\
    % ~~~~~Syrian (n, \%) & 22 (1.21\%) & 4 (0.68\%) & 11 (1.18\%)& \\
    Non-Arab, n (\%) & 1534 (83.9) & 498 (84.8) & 591 (83.1)& 182 (80.9) \\
    % ~~~~~Indian (n, \%) & 760 (41.69\%) & 202 (34.24\%) & 366 (39.27\%) &\\
    % ~~~~~Pakistani (n, \%) & 268 (14.70\%) & 117 (19.83\%) & 165 (17.70\%)&\\
    % ~~~~~Bangladeshi (n, \%) & 210 (11.52\%) & 74 (12.54\%) & 111 (11.91\%)&\\
    % ~~~~~Filipino (n, \%) & 109 (5.98\%) & 44 (7.46\%) & 60 (6.44\%)&\\
    % ~~~~~Nepalese (n, \%) & 92 (5.05\%) & 26 (4.41\%) & 36 (3.86\%) &\\
    %Asian (n, \%) & 1458 (79.98\%)  & 466 (78.98\%) &  747 (80.15\%)  \\
    %CAUROCasian (n, \%) &  11 (0.60\%)  & 1 (0.17\%) & 9 (0.97\%)\\ 
    %Others and unknown (n, \%)  & 55 (3.02\%) & 23 (3.90\%) & 32 (3.43\%)\\ 

    % \textbf{Comborbidities}  & & &\\
    % \midrule
    % Diabetes (n, \%)  & 393 (21.56\%) & 211 (35.76\%) & 238 (25.54\%)\\ 
    % Hypertension (n, \%) & 526 (28.85\%) & 206 (34.92\%) & 270 (28.97\%)\\ 
    % Cancer (n, \%)  &  31 (1.70\%)  & 13 (2.20\%)  & 13 (1.39\%) \\ 
    % % Chronic kidney disease (n, \%) & 57 (3.13\%) & 26 (4.41\%) & 42 (4.51\%)\\ 
    %  \midrule
    
    % Hypertension & 550 (30.1\%)& 213 (36.3\%)&168 (23.6\%)& 71 (31.6\%)\\
    % Diabetes & 427(23.3\%)& 221(37.6\%) & 121 (17.0\%) & 73 (32.4\%) \\
    % Chronic Kidney Disease & 68 (3.7\%)& 30 (5.1\%)& 20 (2.8\%)& 7 (3.1\%)\\
    % Cancer & 30 (1.6\%) & 7(1.2\%) & 12 (1.7\%) & 8(3.6\%) \\

    Critical Care, n (\%)& 332 (18.2) & 83 (14.1) & 63 (8.8) & 26 (11.6) \\
    Mortality, n (\%) & 36 (2.0) &  22 (3.7) &  9 (1.3) &   3 (1.3)\\  
    %Prolonged hospital stay, n (\%)  & 211 (11.5) &  96 (16.4) &    91 (12.7) &   40 (15.7)\\   

    \midrule
    \textbf{Complications}  & & &&\\
    \midrule

    % \textbf{Non-mortal Complications} &&&\\
    % \midrule
     Elevated troponin, n (\%) & 101 (5.5) & 50 (8.5) & 33 (4.6)& 19 (8.4)\\
     ~~~~~  Developed within 24 hours from admission, n (\%) & 47 (2.6) & 36 (6.1) & 20 (2.8) & 5 (2.2) \\  
     ~~~~~ Developed after 24 hours from admission, n (\%) & 54 (3.0) & 14 (2.4) & 13 (1.8) & 14 (6.2) \\  
         \midrule

    Elevated d-dimer, n (\%) & 643 (35.2) & 296 (50.4) & 173 (24.3)& 78 (34.7) \\
   ~~~~~ Developed within 24 hours from admission, n (\%) & 523 (28.6) & 268 (45.7)  &  130 (18.3)& 60 (26.7) \\ 
     ~~~~~ Developed after 24 hours from admission, n (\%) & 120 (6.6) & 28 (4.8) & 43 (6.0) & 18 (8.0) \\

     \midrule
  
    Elevated aminotransferases, n (\%) & 397 (21.7) & 117 (30.2) & 119 (16.7)& 56 (24.9)\\
    ~~~~~   Developed within 24 hours from admission, n (\%) & 287 (15.7) & 133 (22.7) & 72 (10.1) & 35 (15.6)\\
    ~~~~~   Developed after 24 hours from admission, n (\%) & 110 (6.0) & 44 (7.5)  &  47 (6.6) & 21 (9.3)\\
       \midrule
        Elevated interleukin-6, n (\%)  & 245 (13.5) & 126 (21.5) & 65 (9.1)& 28 (12.4)\\
   ~~~~~    Developed within 24 hours from admission, n (\%)  & 57 (3.1) & 49 (8.3)  &  7 (1.0)& 1 (0.4)\\ 
    ~~~~~    Developed after 24 hours from admission, n (\%)  & 188 (10.3) & 77 (13.1)  &  58 (8.2)& 27 (12.0)\\ 
      \midrule

    SBI, n (\%) & 92 (5.0) & 45 (7.7) & 23 (3.2)& 17 (7.6)\\
    ~~~~~   Developed within 24 hours from admission, n (\%) & 1 (0.1) & 3 (0.5)  &  1 (0.1) & 1 (0.4)\\ 
      ~~~~~    Developed after 24 hours from admission, n (\%)  & 91 (5.0) & 42 (7.2)  &  22 (3.1)& 16 (7.1)\\ 
      \midrule  
      AKI, n (\%) & 126 (6.9) & 52 (8.9) & 32 (4.5) & 16 (7.1) \\
     ~~~~~ Developed within 24 hours from admission, n (\%) & 28 (1.5) & 9 (1.5)  &  14 (2.0)& 3 (1.3)\\  
     ~~~~~    Developed after 24 hours from admission, n (\%)  & 98 (5.4) & 43 (7.3)  &  18 (2.5)& 13 (5.8)\\ 
     \midrule
      
    ARDS, n (\%) & 117 (6.4) & 57 (9.7) & 45 (6.3) & 24 (10.7)\\
     ~~~~~  Developed within 24 hours from admission, n (\%) & 61 (3.3) & 26 (4.4)  &  23 (3.2) & 13 (5.8)\\  
     ~~~~~    Developed after 24 hours from admission, n (\%)  & 56 (3.1) & 31 (5.3)  &  22 (3.1)& 11 (4.9)\\

    \bottomrule
    \end{tabular}
}
    \label{tab:charcteristics} 
\end{table}

\begin{table*}[ht!]   
\caption{\small Characteristics of the variables that were used as input features to our models. The mean and interquartile ranges are shown for the demographic features, vital-sign measurements, and laboratory-test results. For the commorbidities and symptoms admission, n denotes the number of patients and \% denotes the percentage of patients per the respective dataset.}   

\centering
    \resizebox{0.9\linewidth}{!}{
    \begin{tabular}{lllll}
    \toprule
    \textbf{Variable, unit} & \textbf{Training set A } & \textbf{Test set A}  & \textbf{Training set B} & \textbf{Test set B} \\
    \midrule
     \textbf{Demographics, mean (IQR)} &&&&\\
    \midrule
    Age & 41.7 (17.0) & 45.5 (18.0) & 39.3 (17.0) & 42.7 (20.0)\\ 
    BMI  & 26.9 (5.2) & 26.7 å(5.7) &  26.5 (5.7) & 27.9 (6.2) \\
    \midrule
    \textbf{Comorbidties, n (\%)} &&&&\\
    \midrule
    Hypertension & 550 (30.1)& 213 (36.3)&168 (23.6)& 71 (31.6)\\
    Diabetes & 427 (23.3)& 221 (37.6) & 121 (17.0) & 73 (32.4) \\
    Chronic kidney disease & 68 (3.7)& 30 (5.1)& 20 (2.8)& 7 (3.1)\\
    Cancer & 30 (1.6) & 7 (1.2) & 12 (1.7) & 8 (3.6) \\
    \midrule
    \textbf{Symptoms at admission, n (\%)} &&&&\\
    \midrule
     Cough & 851 (46.5) & 338 (57.6) & 259 (36.4) & 99 (44.0)\\

    Fever &  28 (1.5) & 20 (3.4) & 3 (0.4) & 3 (1.3)\\
   
    Shortness of breath & 190 (10.4) & 99 (16.9) & 71 (10.0) & 34 (15.1)\\
    Sore throat&  238 (13.0) & 89 (15.2) & 118 (16.6) & 28 (12.4)\\
    Rash & 29 (1.6) & 10 (1.7) & 15 (2.1) & 5 (2.2)\\
    
    \midrule
     \textbf{Laboratory-test results, mean (IQR)} & & & &\\
     \midrule
     Albumin, \textit{g/L}  & 38.4 (11.1) & 35.1 (11.9) & 39.9 (6.0) & 38.6 (7.7)\\
     APTT, \textit{seconds}  & 38.8 (5.6) & 38.1 (6.1) & 30.4 (8.3) & 29.5 (5.7)\\
    Bilirubin, \textit{micromol/L}  & 9.9 (5.5) & 9.3 (6.1) & 9.7 (5.5) & 9.4 (5.0) \\
    Calcium, \textit{mmol/L}  & 2.3 (0.2) & 2.2 (0.2) & 2.3 (0.2) & 2.3 (0.2)\\
    Chloride, \textit{mmol/L}  & 100.9 (4.2) & 99.7 (4.4) & 101.3 (3.3) & 100.7 (4.4)\\ 
     C-reactive protein, \textit{mg/L} & 30.1 (23.4) & 55.7 (74.6) & 22.2 (7.6) & 29.6 (20.8)\\
    Ferritin, \textit{mcg/L} & 579.1 (382.2) & 672.6 (652.4) & 470.9 (288.4) & 624.6 (628.0)\\
     Hematocrit, \textit{L/L}  & 0.4 (0.1) & 0.4 (0.1) & 0.4 (0.1) & 0.4 (0.1)\\
     Hemoglobin, \textit{g/L}  & 141.9 (21.0) & 136.4 (28.0) & 143.4 (22.0) & 141.6 (18.0)\\
     INR, \textit{ratio}  & 1.1 (0.1) & 1.7 (0.1) & 1.0 (0.1) & 1.0 (0.1)\\
     LDH,\textit{ IU/L}  & 260.2 (106.0) & 316.5 (153.0) & 235.3 (68.6) & 263.0 (103.1)\\
    Lymphocytes, \textit{10$^9$/L}  & 1.9 (1.1) & 1.5 (0.9) & 2.0 (1.2) & 1.7 (1.0)\\
    Prothrombin time, \textit{seconds} & 13.9 (1.2) & 13.9 (1.2) & 11.0 (1.6) & 10.8 (1.1)\\
    Procalcitonin, \textit{ng/mL}  & 0.4 (0.1) & 0.8 (0.2) & 0.5 (0.1) & 0.2 (0.1)\\
    Sodium, \textit{mmol/L}  &  138.8 (3.9) & 137.4 (5.5) & 139.4 (3.0) & 138.6 (4.4)\\
    RBC, million/mm$^3$  & 5.1 (0.8) & 5.0 (1.0) & 5.1 (0.78) & 5.1 (0.1)\\
    Urea, \textit{mmol/L} & 4.4 (1.70) & 5.3 (2.1) & 4.2 (1.5) & 4.7 (1.9)\\
    Uric Acid, \textit{mmol/L}   &288.4 (121.0) & 273.6 (100.0) & 295.5 (108.0) & 288.3 (115.5)\\
    Neutrophil, \textit{10$^9$/L} & 4.1 (2.3) & 4.8 (3.0) & 4.2 (2.5) & 4.5 (2.4)\\
    \midrule
     \textbf{Vital-sign measurements, mean (IQR)} & & &\\
     \midrule
     Systolic Blood Pressure, \textit{mmHg}  & 126.3 (15.0) & 126.8 (16.0) & 128.8 (15.5) & 128.2 (15.7)\\
     Diastolic Blood Pressure, \textit{mmHg}  & 77.5 (9.8) & 76.9 (9.9) & 77.9 (10.3) & 77.5 (10.7)\\
     Respiratory Rate, \textit{breaths per minute}  & 18.9 (1.0) & 20.2 (2.5) & 18.1 (0.7) & 18.7 (0.8)\\
     Peripheral Pulse Rate, \textit{beats per minute} & 82.6 (11.5) & 85.4 (11.6) & 81.7 (13.4) & 82.5 (12.5)\\ 
     Oxygen Saturation, \textit{\%}   & 98.4 (1.6) & 97.5 (2.1) & 98.5 (1.0) & 98.2 (1.4)\\ 
     Temperature Auxiliary, \textit{$^{\circ}$C}  & 36.9 (0.4) & 37.0 (0.7) & 36.9 (0.4) & 37.1 (0.6)\\
     Glasgow Coma Score  & 14.8 (0.0) & 15.0 (0.0) & 14.8 (0.0) & 14.8 (0.0)\\
    \bottomrule
    \end{tabular}}
    \label{tab:distributions} 
\end{table*}

%word count goal: 1000 words 
A total of 3,352 encounters were included in the study and the characteristics of the characteristics of the final data splits are presented in Table~\ref{tab:charcteristics}.  Across all the data splits, the mean age ranges between 39.3 and 45.5 years and the proportion of males ranges between 84.8\% to 88.9\%. The mortality rate was also less than 4\% across all data splits, ranging between 1.3\% and 3.7\%.  The most prevalent complication across all datasets was elevated d-dimer, although most patients mainly exhibited elevated d-dimer during the first 24 hours of admission. Elevated interleukin-6 was the most prevalent complication developed after 24 hours of admission across all datasets. The incidence of the complications developed after 24 hours were higher in the test sets than in their respective training sets, except for elevated troponin and d-dimer which were higher in training set A (3.0\% and 6.6\%, respectively) than in test set A (2.4\% and 4.8\%, respectively). The distributions of the vital signs and laboratory-test results, in terms of of the mean and interquartile ranges, are shown in Table~\ref{tab:distributions}.

%This suggests that patients who were admitted in late April were more severe than those who were admitted earlier in April.

%The high proportion of males in our datasets is explained by the influx of male migrant workers in Abu Dhabi.  

%both training set A and B and testing set A and B were elevated D Dimer ( 23.5\% - 50.4\%), followed by elevated aminotransferases (16.7\% - 30.2\% ), elevated Interleukin-6 (9.1\% - 21.5\% ), ARDS (6.3\% - 10.7\% ), AKI (4.5\% - 8.9\% ), elevated cardiac troponin (4.6\% - 8.5\% ) and secondary bacterial infection (3.2\% - 7.7\% ). The incidence of complications is generally higher in the test sets than their respective training sets, suggesting that later admissions in April were more severe than earlier ones. 

\begin{figure}[ht!]
    \centering
     \begin{tabular}{c c c c}
    \includegraphics[width=0.22\linewidth]{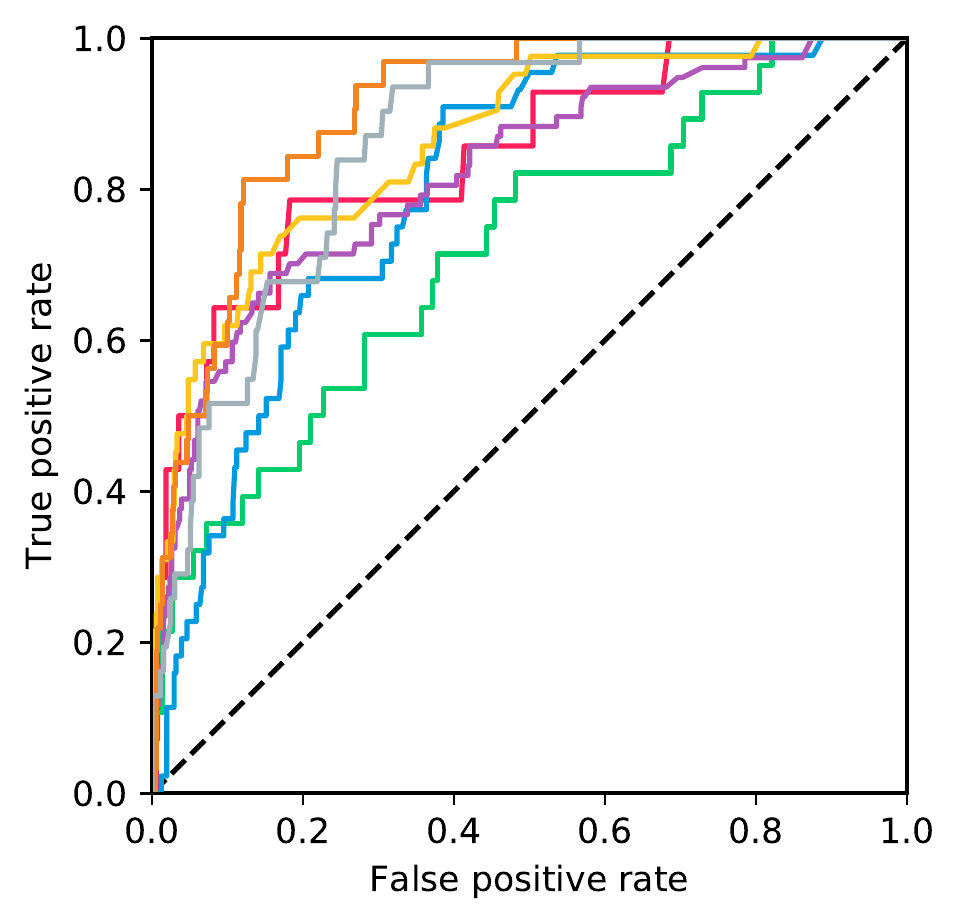} &
    \includegraphics[width=0.22\linewidth]{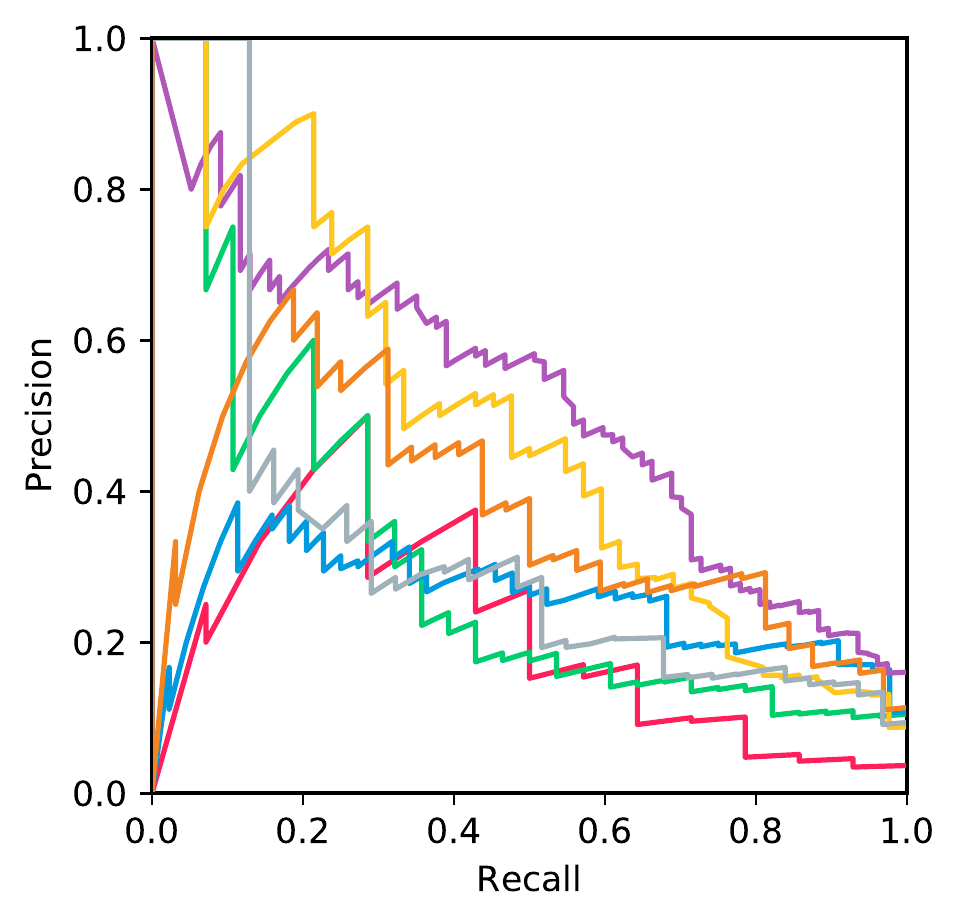} &
    \includegraphics[width=0.22\linewidth]{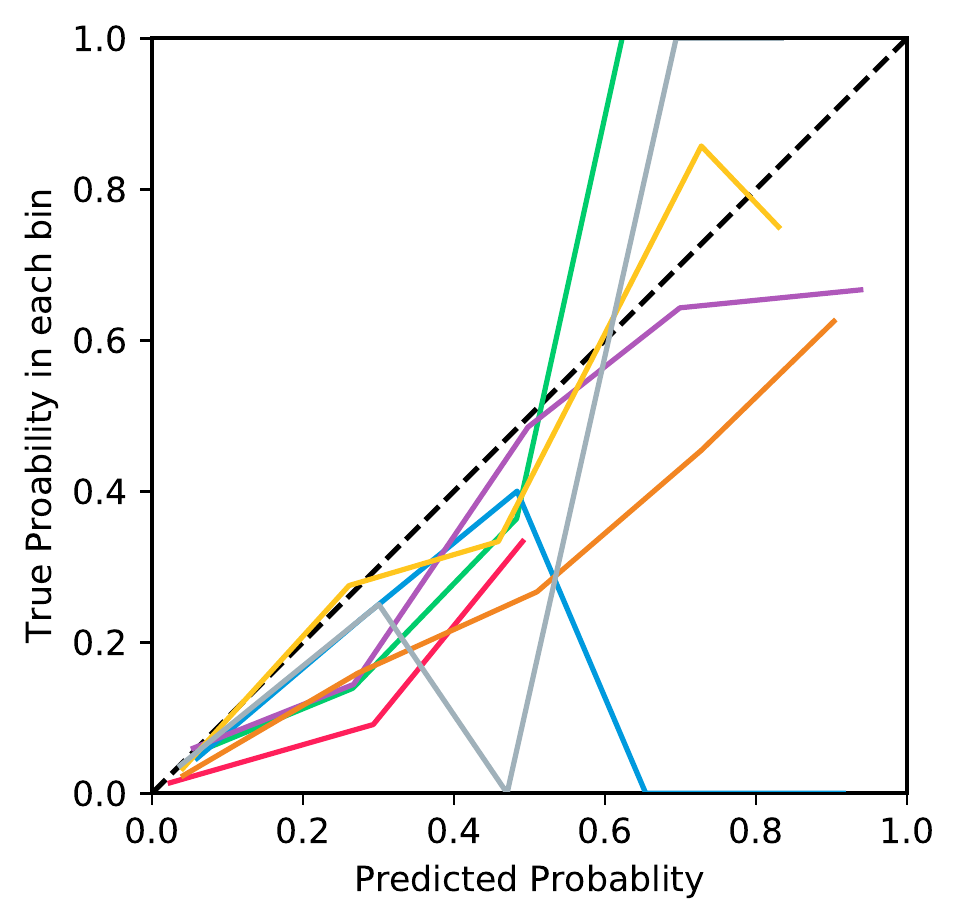} & 
     \includegraphics[width=0.12\linewidth]{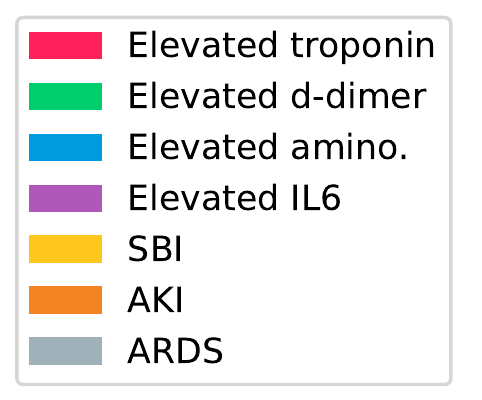}
    \\
    \includegraphics[width=0.22\linewidth]{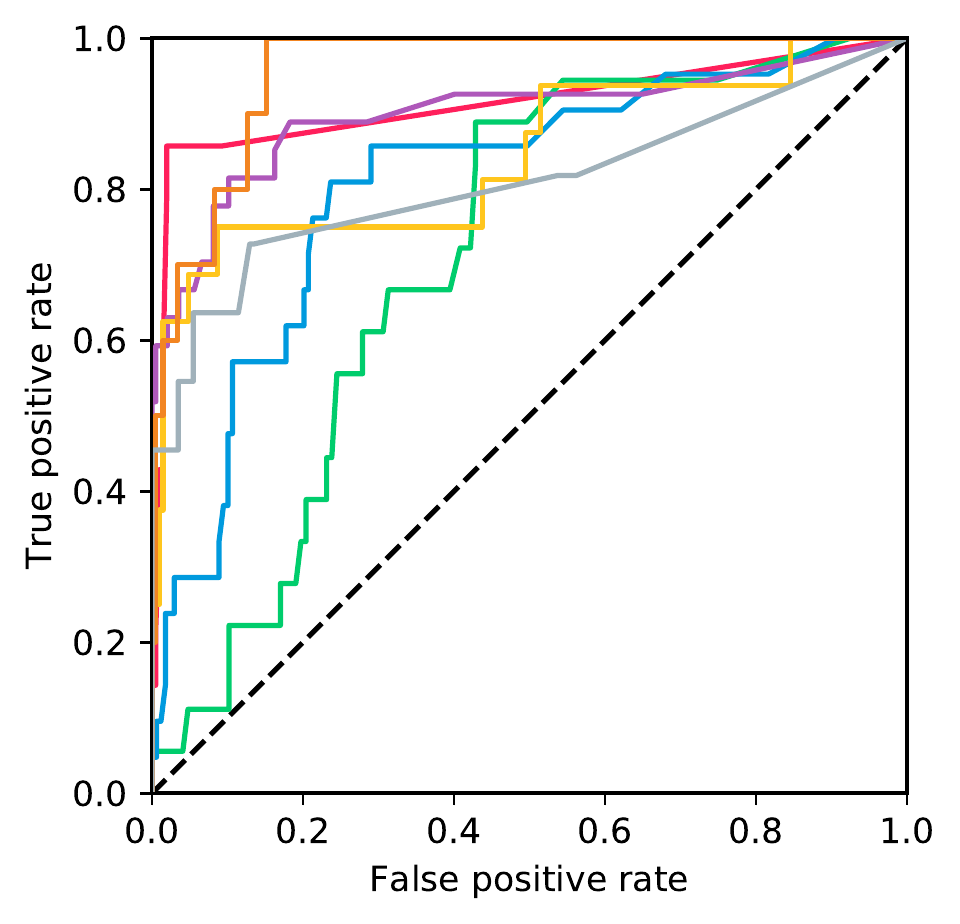}&
    \includegraphics[width=0.22\linewidth]{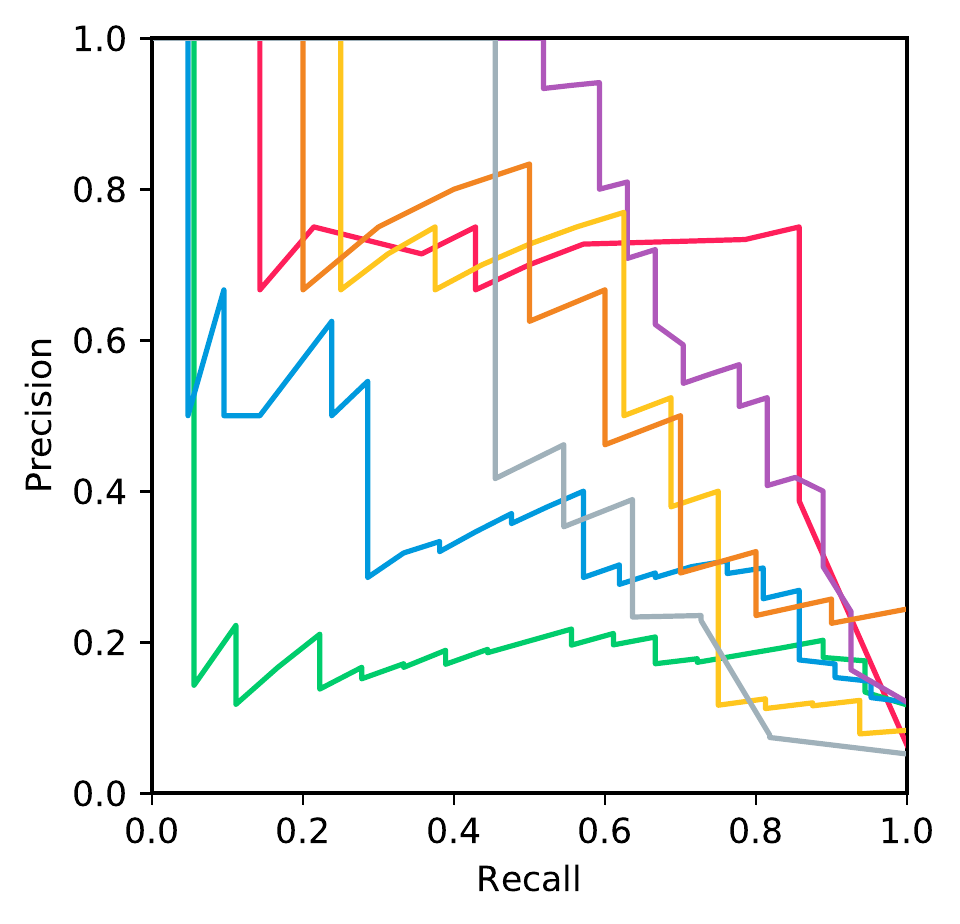}
    & 
    \includegraphics[width=0.22\linewidth]{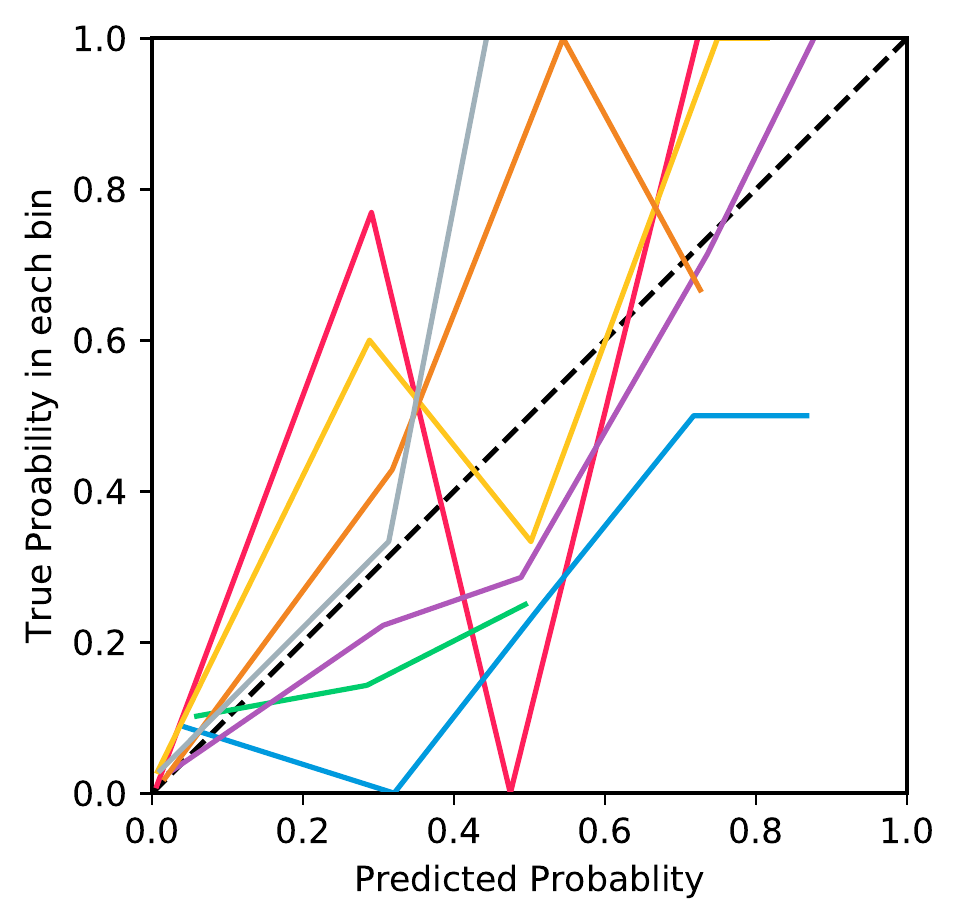}
    & \\
    (a) & (b) & (c) &\\
    \end{tabular}
    \vspace{-2mm}
    \caption{\small The (a) ROC curves, (b) PRC curves, and (c) calibration curves are shown for all model ensembles evaluated on test set A (top) and test set B (bottom). The color legend for all figures is shown on the right. The numerical values for the AUROC, AUPRC, calibration slopes and intercepts can be found in Table \ref{performance}.}
    \label{fig:performance-plots}
\end{figure}

The performance of the models selected by our system across the two test sets in terms of the AUROC and AUPRC are shown in Table~\ref{performance}. The ROC, PRC, and reliability plots are also visualized in Figure~\ref{fig:performance-plots}. Across both test sets, our data-driven approach achieved good accuracy ($>$0.80 AUROC) for most complications. In test set A, AKI was the best discriminated endpoint at 24 hours from admission, with 0.905 AUROC (95\% CI 0.861, 0.946). This is followed by ARDS (0.864 AUROC), SBI (0.862 AUROC), elevated troponin (0.843 AUROC), elevated interleukin-6 (0.820 AUROC), and elevated aminotransferases (0.801 AUROC). The complication with the worst discrimination was elevated d-dimer (0.717 AUROC). In test set B, AKI was also the best discriminated endpoint with 0.958 AUROC (95\% CI 0.913, 0.994), followed by elevated troponin (0.913 AUROC), and elevated interleukin-6 (0.899 AUROC). Similar to test set A, elevated d-dimer was the worst discriminated endpoint (0.714 AUROC). 
We also observe that LGBM was selected as the best performing model on the validation sets for most complications, as shown in Supplementary Section \ref{apd:perf}. LR was selected for AKI in both datasets, for elevated d-dimer in dataset A, and for SBI in dataset B, highlighting its predictive power despite its simplicity compared to the other machine learning models.

\begin{figure}[ht!]

    \centering
   \resizebox{0.9\linewidth}{!}{
    \begin{tabular}{c c }
    \includegraphics[width=0.59\textwidth]{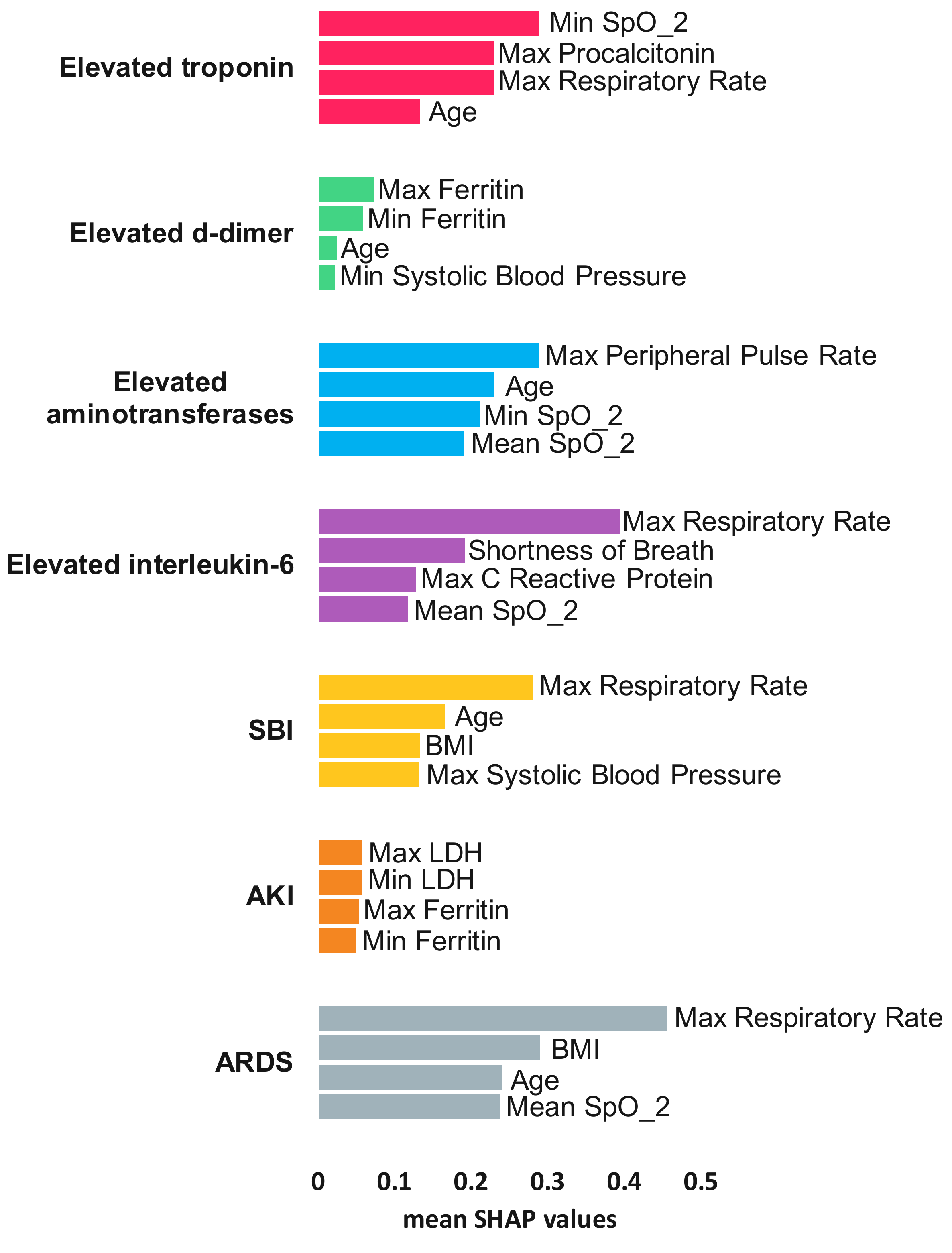} &
    \includegraphics[width=0.6\textwidth]{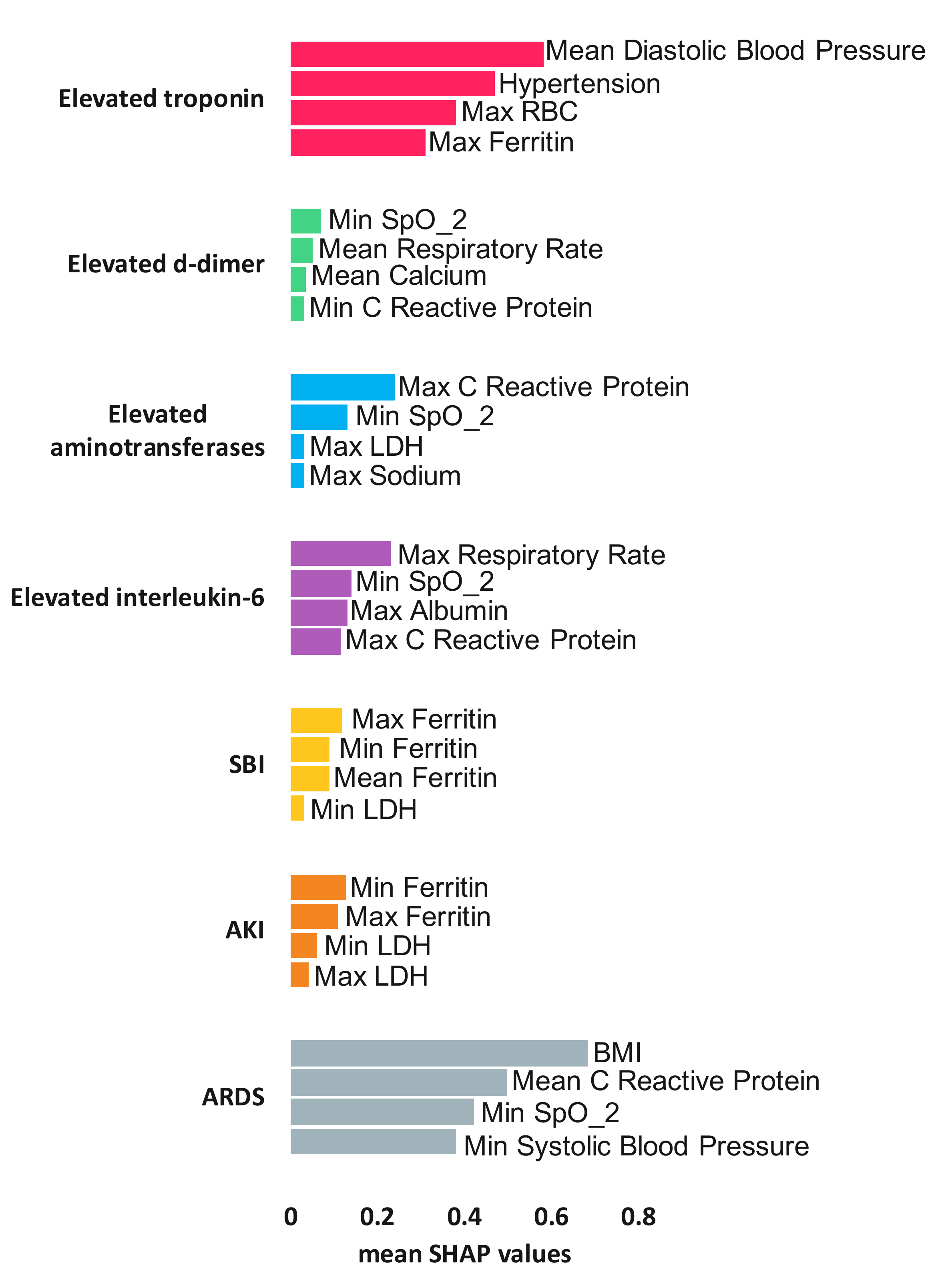} \\
    (a) & (b) \\
    \end{tabular}

}
    \vspace{-2mm}
    \caption{\small The four most important features are shown for each complication in (a) test set A and (b) test set B. Feature importance was computed using the average SHAP values of the six models per ensemble.}
    \label{fig:feature-importance}
\end{figure}

The top four important features for each complication are shown in Figure~\ref{fig:feature-importance} across the two test sets. In test set A, age was among the top predictive features for all the complications except for elevated interleukin-6 and AKI. In test set B, C-reactive protein was  among the top predictive features for predicting elevated aminotransferases, elevated d-dimer, elevated interleukin-6, and ARDS. Other features such as ferritin and LDH, and BMI, were among the top predictive features for several complications across both sets, specifically for AKI and ARDS, respectively. 

We also visualize the timeline of for two patients in Figure~\ref{fig:examplepatients}, along with the predictions of our system. In Figure~\ref{fig:examplepatients}(a), the patient shown developed all seven complications during their hospital stay of 43 days. This highlights the importance of predicting all complications simultaneously, especially for patients who may develop more than one complication. In Figure~\ref{fig:examplepatients}(b), the patient did not develop any complications during their hospital stay of two days. To compare both patients, the system's predictions for patient (a) were relatively higher than those for patient (b). For example, the AKI predictions were 0.54 and 0.03, respectively, despite the fact that patient (a) developed AKI at around 20 days from admission. This demonstrates the value of our system in predicting the risk of developing complications early during the patient's stay.

\begin{table}[ht!]  
\centering
 \caption{\small Performance evaluation of the best performing models on test sets A \& B, which were selected based on the average AUROC performance on the validation sets, as shown in Supplementary Section \ref{apd:perf}. Model type indicates the type of the base learners within the final selected ensemble. All the metrics were computed using bootstrapping with 1,000 iterations \cite{diciccio1996bootstrap}.}
 \resizebox{0.8\linewidth}{!}{
\begin{tabular}{llcc}
\toprule
\textbf{Complication}  &  \textbf{Result} & {\textbf{Test Set A}}&\textbf{Test Set B} \\
\midrule
    & Model type & LGBM & LGBM \\
   & AUROC &  0.843 (0.720, 0.945) & 0.913 (0.788, 0.994) \\
    & AUPRC & 0.226 (0.106, 0.499)& 0.674 (0.405, 0.898) \\
    \textbf{Elevated troponin} & Calibration Slope & 0.661 (0.134, 1.264) & 1.029 (0.142, 2.536)\\
        & Calibration Intercept & -0.032 (-0.101, 0.029) &  0.060 (-0.027, 0.257)\\
        
      \midrule
   & Model type & LR & LGBM \\
    & AUROC & 0.717 (0.618, 0.816) & 0.714 (0.612, 0.810) \\
    & AUPRC & 0.315 (0.167, 0.494) & 0.235 (0.118, 0.397)\\
    \textbf{Elevated d-dimer} & Calibration Slope&  1.592 (0.460, 1.841) & 0.338 (-0.33, 1.481)\\
    & Calibration Intercept& -0.187 (-0.241, 0.023) & 0.071 (-0.102, 0.206) \\
          \midrule
        & Model type & LGBM & LGBM  \\
     & AUROC &  0.801 (0.741, 0.858) & 0.808 (0.699, 0.894)\\
    & AUPRC &  0.261 (0.176, 0.391) & 0.396 (0.229, 0.604) \\
    \textbf{Elevated aminotransferases } & Calibration Slope & -0.145 (-0.193, 0.159) & 0.628 (0.172, 1.135)\\
    & Calibration Intercept& 0.205 (0.110, 0.254) &0.042 (-0.110, 0.186)\\
    
      \midrule
      & Model type & LGBM & LGBM \\
     & AUROC & 0.820 (0.760, 0.872)& 0.899 (0.810, 0.971)\\
    & AUPRC &0.514 (0.403, 0.635)&  0.776 (0.623, 0.900)\\

    \textbf{Elevated interleukin-6 } & Calibration Slope&  0.777  (0.540, 0.980) & 1.120 (0.879, 1.299)\\
    & Calibration Intercept& 0.018 (-0.046, 0.094) &  -0.094 (-0.193, 0.034)\\
    
      \midrule
      & Model type & LGBM & LR \\
     & AUROC &   0.862 (0.802, 0.920) & 0.843 (0.721, 0.960)\\
    & AUPRC &   0.486 (0.339, 0.645) & 0.612 (0.384, 0.847) \\
    \textbf{SBI} & Calibration Slope&  0.977 (0.566, 1.298) &1.583 (0.846, 1.865)\\
    & Calibration Intercept& -0.021 (-0.126, 0.095)& -0.075 (-0.155, 0.051)\\
    
     \midrule

        & Model type & LR & LR \\
     & AUROC &0.905 (0.861, 0.946) & 0.958 (0.913, 0.994) \\
    &  AUPRC & 0.377 (0.238, 0.574)& 0.637 (0.344, 0.901)\\
    \textbf{AKI} & Calibration Slope&  0.683 (0.313, 1.032) &  1.127 (-0.022, 1.808)\\
    & Calibration Intercept& -0.030 (-0.129, 0.079)&0.076 (-0.106, 0.216)\\
    
    \midrule
    & Model type & LGBM & LGBM \\
     & AUROC &   0.864 (0.809, 0.910) & 0.808 (0.621, 0.972)\\
    & AUPRC & 0.340 (0.202, 0.496) & 0.570 (0.273, 0.842) \\
    \textbf{ARDS} & Calibration Slope& 1.346 (1.082, 1.419) & 2.023 (0.486, 2.18)\\
    & Calibration Intercept&  -0.171 (-0.222, -0.071) & -0.063 (-0.136, 0.032)\\
    \bottomrule
\end{tabular}}
\label{performance}
\end{table}

\section{Discussion}

\begin{figure}[t!]
    \centering
    \begin{tabular}{c }
    \includegraphics[width=0.95\textwidth]{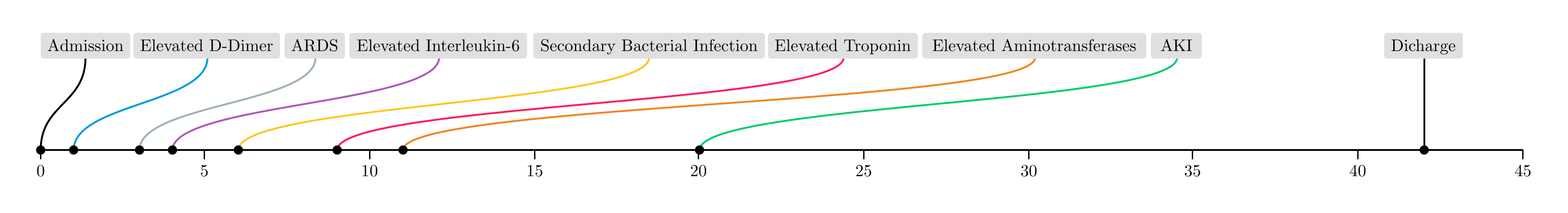}
    \\
    (a) \\
    \includegraphics[width=0.95\textwidth]{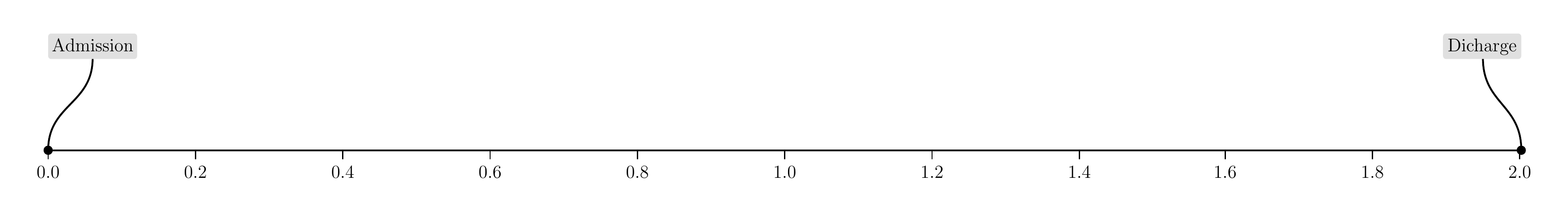}\\
    (b) \\
    \end{tabular}
    \vspace{-2mm}
    \caption{\small Timeline showing the development of complications with respect to number of days from admission (x-axis) for two sample patients. (a) For $[y\textsuperscript{El. troponin}, y\textsuperscript{El. d-dimer}, y\textsuperscript{El. aminotransferases},   y\textsuperscript{El. interleukin-6},  y\textsuperscript{SBI}, y\textsuperscript{AKI}, y\textsuperscript{ARDS}]$, our system predictions (multiplied by a 100 to obtain percentages) were [71\%, NA, 60\%, 87\%, 77\%, 54\%, 44\%], where NA denotes that the complication had already occurred during the first 24 hours, hence it was not predicted. (b) This patient did not develop and complications and our model predictions were [2\%, 7\%, 1\%, 16\%, 1\%, 3\%, 2\%].}
    \label{fig:examplepatients}
\end{figure}

In this study, we developed a predictive system of commonly occurring complications among COVID-19 patients to support patient triage. During validation, the system was assessed for performance and calibration. To the best of our knowledge, this is one of the few machine learning studies that predict non-mortal complications secondary to COVID-19 and the first to demonstrate a system that predicts the risk of such complications simultaneously. The system achieves a good performance across all complications, for example, reaching above 0.9 AUROC for AKI across two independent datasets. This study has several strengths and limitations. 

% (1) discuss the dataset 
One of the main strengths is that we used multicentre data collected from 18 facilities across several regions in Abu Dhabi, UAE. COVID-19 treatment is free for all patients, hence there were no obvious gaps in terms of access to healthcare services in our dataset. Our dataset is diverse since Abu Dhabi is residence for more than 200 nationalities, of which only 19.0\% of the population only is Emirati. Those characteristics of the dataset make our findings relevant to a global audience. This is also the first data-driven study to present the population in the UAE and one of few studies with large sample sizes (3,352 COVID-19 patient encounters) among COVID-19 related studies, while most previous studies have focused on European or Chinese patient cohorts. Despite the diversity of the dataset, one limitation is that we did not perform validation on a patient cohort external to the UAE. Compared to other international patient cohorts, our patient cohort is relatively younger, with a lower  overall mortality rate, suggesting that our system needs to be further validated on populations with different demographic distributions ~\cite{2020FeiTingRonghui,elevated_troponin1,2020GoyalChoi, ards1}. Our data-driven approach and open-access code can be easily adapted for such purposes. 

% (2) discuss the choice of complications 
Several studies reported worse prognosis among COVID-19 infected patients who had multi-organ failure, severe inflammatory response, and other hematological complications \cite{ddimer2, kindey2,elevated_troponin1,amino2}. Most existing studies focus on predicting the mortality endpoint~\cite{critical}. The low mortality rate in our dataset strongly discouraged the development of a mortality risk prediction score, as small sample sizes may lead to biased models~\cite{critical}. Our work was motivated by predicting the precursors of such severe adverse events, as identified by the World Health Organization~\cite{WHO_lancet}. We identified and predicted seven complications indicative of patient severity in order to avoid worse patient outcomes. The prevalence of the predicted complications ranged between 2\%-10\% and 2\%-13\% in our training and test sets, respectively. This high class imbalance is reflected in the AUPRC results. Since most of those tasks have not been investigated thoroughly before, our results introduce new benchmarks to evaluate other competing models. Future work should also investigate the use of multi-label deep learning classifiers, while accounting for the exclusion criteria during training.

An important aspect of this study is that the labeling criteria relies on renowned clinical standards and hospital-acquired data to identify the exact time of the occurrence of such complications. In collaboration with the clinical experts, this approach was considered more reliable than relying on International Classification of Disease (ICD) codes, since ICD codes are generally used for billing purposes and their derivation may vary across facilities, especially during a pandemic. One limitation of the labeling procedure is that it could miss patients for whom the data used in identifying a particular complication was not collected. However, this issue is more closely related to data collection practices at institutions and clinical data is often not missing at random. We also avoided label leakage by ensuring that there is no overlap between the set of input features and the features used to identify complications. 
%We hope that with early predictions of these complications, early clinical decisions can be taken to prevent adverse outcomes such as myocardial infarction, cytokine storm, coagulation derangement, liver injury and the need for intubation. 

%The factors that lead to mortality in COVID-19 are not completely understood and our findings can contribute to the clinical understanding of COVID-19 related deterioration. 

% (3) features - importance/cost
The feature importance analysis revealed that age, oxygen saturation, and respiratory rate are highly predictive of several complications. Since COVID-19 is predominantly a pulmonary illness, it was not surprising that oxygen saturation and respiratory rate ranked among the highest predictive features. Such features are routinely collected at hospitals and do not incur any additional data collection costs. We also identified C-reactive protein, ferritin, LDH, procalcitonin, systolic blood pressure, and diastolic blood pressure as markers for severity among COVID-19 patients, which is aligned with clinical literature~\cite{features_PCT, ran2020blood}. This analysis demonstrates that our system's learning is clinically meaningful and relevant.

% (4) calibration 
We assessed our models' calibration by reporting the calibration slopes and intercepts with confidence intervals and visualizing the calibration curves. Sufficiently large datasets are usually needed to produce stable calibration curves at model validation stage. Despite the size of our dataset, we found that reporting the calibration slopes and intercepts would provide a concise summary of potential problems with our system's risk calibration, to avoid harmful decision-making ~\cite{van2016calibration}. Overall, our results show that our ensemble models were adequately calibrated across all complications, as shown in Table \ref{performance} and Figure~\ref{fig:performance-plots}(c). This is also reflected in the sample patient timelines shown in Figure~\ref{fig:examplepatients}, where the predicted risks for the patient who experienced the complications were relatively higher than those predicted for the patient who did not experience any complications. Limiting factors to perfect calibration are the small dataset size and the fact that the ensemble prediction consists of an average of the predictions of the individually calibrated models. Further work should investigate how to improve the calibration of ensemble models.

%risks of developing the target complications  

% (5) potential clinical use and implications 
Our data-driven approach and results highlight the promise of machine learning in predicting the risk of complications among COVID-19 patients. The proposed approach performs well when applied to two independent multicentre training and test sets in the UAE. The system can be easily implemented in practice due to several factors. First, the input features that our system uses are routinely collected by hospitals that accommodate COVID-19 patients as recommended by WHO. Second, training the machine learning models within our system does not require high computational resources. Finally, through feature importance analysis, our system can offer interpretability, and is also fully automated as it does not require any manual interventions. To conclude, we propose a clinically applicable prognostic system that predicts non-mortal complications among COVID-19 patients. Our system can serve as a guide to anticipate the course of COVID-19 patients and to help initiate more targeted and complication-specific decision-making on treatment and triage.

\section{Contributors}
GOG, BA, and KWY managed and analyzed the data. FS and IQ extracted, anonymized, and provided the dataset for analysis. GOG, KWY, and NH developed and maintained the experimental codebase. FAK, SAJ, MAS, RA, and MAH provided clinical expertise. WZ, FS, MAH and FES designed the study. MAH and FES supervised the work. GOG, BA, KWY, and FES wrote the manuscript. All authors interpreted the results and revised and approved the final manuscript.

\section{Data sharing}
We are unable to share the full dataset used in this study due to restrictions by the data provider. However, to allow for reproducibility and benchmarking on our dataset, we are sharing test set B (n=225), the trained models, and the source code online at \url{ https://github.com/nyuad-cai/COVID19Complications}. 
 
\section{Acknowledgments} 
We would like to thank NYU Abu Dhabi for the generous funding. We would also like to thank Waqqas Zia and Benoit Marchand from the Dalma team at New York University Abu Dhabi for supporting data management and access to computational resources. This study was supported through the data resources and staff expertise provided by Abu Dhabi Health Services. 

\bibliographystyle{unsrt}
\bibliography{main}

\begin{thebibliography}{10}

\bibitem{dong2020interactive}
Ensheng Dong, Hongru Du, and Lauren Gardner.
\newblock An interactive web-based dashboard to track covid-19 in real time.
\newblock {\em The Lancet infectious diseases}, 20(5):533--534, 2020.

\bibitem{critical}
Laure Wynants, Ben Van~Calster, Gary~S Collins, Richard~D Riley, Georg Heinze,
  Ewoud Schuit, Marc~MJ Bonten, Darren~L Dahly, Johanna~AA Damen, Thomas~PA
  Debray, et~al.
\newblock Prediction models for diagnosis and prognosis of covid-19: systematic
  review and critical appraisal.
\newblock {\em bmj}, 369, 2020.

\bibitem{WHO_lancet}
John~C Marshall, Srinivas Murthy, Janet Diaz, Neil Adhikari, Derek~C Angus,
  Yaseen~M Arabi, Kenneth Baillie, Michael Bauer, Scott Berry, Bronagh
  Blackwood, et~al.
\newblock A minimal common outcome measure set for covid-19 clinical research.
\newblock {\em The Lancet Infectious Diseases}, 2020.

\bibitem{2020FeiTingRonghui}
Fei Zhou, Ting Yu, Ronghui Du, Guohui Fan, Ying Liu, Zhibo Liu, Jie Xiang,
  Yeming Wang, Bin Song, Xiaoying Gu, Lulu Guan, Yuan Wei, Hui Li, Xudong Wu,
  Jiuyang Xu, Shengjin Tu, Yi~Zhang, Hua Chen, and Bin Cao.
\newblock Clinical course and risk factors for mortality of adult inpatients
  with covid-19 in wuhan, china: a retrospective cohort study.
\newblock {\em The Lancet}, 395(10229):1054 -- 1062, 2020.

\bibitem{ddimer2}
Jeffrey~S Berger, Dennis Kunichoff, Samrachana Adhikari, Tania Ahuja, Nancy
  Amoroso, Yindalon Aphinyanaphongs, Meng Cao, Ronald Goldenberg, Alexander
  Hindenburg, James Horowitz, et~al.
\newblock Prevalence and outcomes of d-dimer elevation in hospitalized patients
  with covid-19.
\newblock {\em Arteriosclerosis, thrombosis, and vascular biology},
  40(10):2539--2547, 2020.

\bibitem{kindey2}
Jamie~S. Hirsch, Jia~H. Ng, Daniel~W. Ross, Purva Sharma, Hitesh~H. Shah,
  Richard~L. Barnett, Azzour~D. Hazzan, Steven Fishbane, Kenar~D. Jhaveri,
  Mersema Abate, and et~al.
\newblock Acute kidney injury in patients hospitalized with covid-19.
\newblock {\em Kidney International}, 98(1):209–218, 2020.

\bibitem{elevated_troponin1}
Carlo~Mario Lombardi, Valentina Carubelli, Annamaria Iorio, Riccardo~M
  Inciardi, Antonio Bellasi, Claudia Canale, Rita Camporotondo, Francesco
  Catagnano, Laura~A Dalla~Vecchia, Stefano Giovinazzo, et~al.
\newblock Association of troponin levels with mortality in italian patients
  hospitalized with coronavirus disease 2019: results of a multicenter study.
\newblock {\em JAMA cardiology}, 2020.

\bibitem{MI_china}
Bo~Li, Jing Yang, Faming Zhao, Lili Zhi, Xiqian Wang, Lin Liu, Zhaohui Bi, and
  Yunhe Zhao.
\newblock Prevalence and impact of cardiovascular metabolic diseases on
  covid-19 in china.
\newblock {\em Clinical Research in Cardiology}, 109(5):531–538, 2020.

\bibitem{ddimer1}
Yumeng Yao, Jiatian Cao, Qingqing Wang, Qingfeng Shi, Kai Liu, Zhe Luo, Xiang
  Chen, Sisi Chen, Kaihuan Yu, Zheyong Huang, et~al.
\newblock D-dimer as a biomarker for disease severity and mortality in covid-19
  patients: a case control study.
\newblock {\em Journal of intensive care}, 8(1):1--11, 2020.

\bibitem{amino2}
Qi~Wang, Hong Zhao, Li-Gai Liu, Yan-Bin Wang, Ting Zhang, Ming-Hui Li, Yan-Li
  Xu, Gui-Ju Gao, Hao-Feng Xiong, Ying Fan, et~al.
\newblock Pattern of liver injury in adult patients with covid-19: a
  retrospective analysis of 105 patients.
\newblock {\em Military Medical Research}, 7(1):1--8, 2020.

\bibitem{amino1}
Safiya Richardson, Jamie~S. Hirsch, Mangala Narasimhan, James~M. Crawford,
  Thomas McGinn, Karina~W. Davidson, , and the Northwell COVID-19
  Research~Consortium.
\newblock {Presenting Characteristics, Comorbidities, and Outcomes Among 5700
  Patients Hospitalized With COVID-19 in the New York City Area}.
\newblock {\em JAMA}, 323(20):2052--2059, 05 2020.

\bibitem{elevated_il6}
Zeming Liu, Jinpeng Li, Danyang Chen, Rongfen Gao, Wen Zeng, Sichao Chen, Yihui
  Huang, Jianglong Huang, Wei Long, Man Li, Liang Guo, Xinghuan Wang, and
  Xiaohui Wu.
\newblock Dynamic interleukin-6 level changes as a prognostic indicator in
  patients with covid-19.
\newblock {\em Frontiers in Pharmacology}, 11:1093, 2020.

\bibitem{2020GoyalChoi}
Parag Goyal, Justin~J. Choi, Laura~C. Pinheiro, Edward~J. Schenck, Ruijun Chen,
  Assem Jabri, Michael~J. Satlin, Thomas~R. Campion, Musarrat Nahid, Joanna~B.
  Ringel, Katherine~L. Hoffman, Mark~N. Alshak, Han~A. Li, Graham~T. Wehmeyer,
  Mangala Rajan, Evgeniya Reshetnyak, Nathaniel Hupert, Evelyn~M. Horn,
  Fernando~J. Martinez, Roy~M. Gulick, and Monika~M. Safford.
\newblock Clinical characteristics of covid-19 in new york city.
\newblock {\em New England Journal of Medicine}, 382(24):2372--2374, 2020.

\bibitem{ards1}
Kyung~Soo Hong, Kwan~Ho Lee, Jin~Hong Chung, Kyeong-Cheol Shin, Eun~Young Choi,
  Hyun~Jung Jin, Jong~Geol Jang, Wonhwa Lee, and June~Hong Ahn.
\newblock Clinical features and outcomes of 98 patients hospitalized with
  sars-cov-2 infection in daegu, south korea: a brief descriptive study.
\newblock {\em Yonsei Medical Journal}, 61(5):431, 2020.

\bibitem{guan2020clinical}
Wei-jie Guan, Zheng-yi Ni, Yu~Hu, Wen-hua Liang, Chun-quan Ou, Jian-xing He,
  Lei Liu, Hong Shan, Chun-liang Lei, David~SC Hui, et~al.
\newblock Clinical characteristics of coronavirus disease 2019 in china.
\newblock {\em New England journal of medicine}, 382(18):1708--1720, 2020.

\bibitem{collins2015transparent}
Gary~S Collins, Johannes~B Reitsma, Douglas~G Altman, and Karel~GM Moons.
\newblock Transparent reporting of a multivariable prediction model for
  individual prognosis or diagnosis (tripod) the tripod statement.
\newblock {\em Circulation}, 131(2):211--219, 2015.

\bibitem{khwaja2012kdigo}
Arif Khwaja.
\newblock Kdigo clinical practice guidelines for acute kidney injury.
\newblock {\em Nephron Clinical Practice}, 120(4):c179--c184, 2012.

\bibitem{2012ARDS_Berlin}
The ARDS Definition~Task Force.
\newblock {Acute Respiratory Distress Syndrome: The Berlin Definition}.
\newblock {\em JAMA}, 307(23):2526--2533, 06 2012.

\bibitem{troponins}
Zhivko Zhelev, Christopher Hyde, Emily Youngman, Morwenna Rogers, Simon
  Fleming, Toby Slade, Helen Coelho, Tracey Jones-Hughes, and Vasilis Nikolaou.
\newblock Diagnostic accuracy of single baseline measurement of elecsys
  troponin t high-sensitive assay for diagnosis of acute myocardial infarction
  in emergency department: systematic review and meta-analysis.
\newblock {\em Bmj}, 350:h15, 2015.

\bibitem{features_PCT}
Ian Huang, Raymond Pranata, Michael~Anthonius Lim, Amaylia Oehadian, and Bachti
  Alisjahbana.
\newblock C-reactive protein, procalcitonin, d-dimer, and ferritin in severe
  coronavirus disease-2019: a meta-analysis.
\newblock {\em Therapeutic advances in respiratory disease},
  14:1753466620937175, 2020.

\bibitem{bergstra2012random}
James Bergstra and Yoshua Bengio.
\newblock Random search for hyper-parameter optimization.
\newblock {\em The Journal of Machine Learning Research}, 13(1):281--305, 2012.

\bibitem{van2016calibration}
Ben Van~Calster, Daan Nieboer, Yvonne Vergouwe, Bavo De~Cock, Michael~J
  Pencina, and Ewout~W Steyerberg.
\newblock A calibration hierarchy for risk models was defined: from utopia to
  empirical data.
\newblock {\em Journal of clinical epidemiology}, 74:167--176, 2016.

\bibitem{lundberg2020local}
Scott~M Lundberg, Gabriel Erion, Hugh Chen, Alex DeGrave, Jordan~M Prutkin,
  Bala Nair, Ronit Katz, Jonathan Himmelfarb, Nisha Bansal, and Su-In Lee.
\newblock From local explanations to global understanding with explainable ai
  for trees.
\newblock {\em Nature machine intelligence}, 2(1):2522--5839, 2020.

\bibitem{LGBM}
Guolin Ke, Qi~Meng, Thomas Finley, Taifeng Wang, Wei Chen, Weidong Ma, Qiwei
  Ye, and Tie-Yan Liu.
\newblock Lightgbm: A highly efficient gradient boosting decision tree.
\newblock In I.~Guyon, U.~V. Luxburg, S.~Bengio, H.~Wallach, R.~Fergus,
  S.~Vishwanathan, and R.~Garnett, editors, {\em Advances in Neural Information
  Processing Systems 30}, pages 3146--3154. Curran Associates, Inc., 2017.

\bibitem{diciccio1996bootstrap}
Thomas~J DiCiccio and Bradley Efron.
\newblock Bootstrap confidence intervals.
\newblock {\em Statistical science}, pages 189--212, 1996.

\bibitem{ran2020blood}
Jinjun Ran, Ying Song, Zian Zhuang, Lefei Han, Shi Zhao, Peihua Cao, Yan Geng,
  Lin Xu, Jing Qin, Daihai He, et~al.
\newblock Blood pressure control and adverse outcomes of covid-19 infection in
  patients with concomitant hypertension in wuhan, china.
\newblock {\em Hypertension Research}, pages 1--10, 2020.

\bibitem{herasevich2009validation}
Vitaly Herasevich, Murat Yilmaz, Hasrat Khan, Rolf~D Hubmayr, and Ognjen Gajic.
\newblock Validation of an electronic surveillance system for acute lung
  injury.
\newblock {\em Intensive care medicine}, 35(6):1018--1023, 2009.

\bibitem{2009AzzamHelen}
Helen~C. Azzam, Satjeet~S. Khalsa, Richard Urbani, Chirag~V. Shah, Jason~D.
  Christie, Paul~N. Lanken, and Barry~D. Fuchs.
\newblock {Validation Study of an Automated Electronic Acute Lung Injury
  Screening Tool}.
\newblock {\em Journal of the American Medical Informatics Association},
  16(4):503--508, 07 2009.

\end{thebibliography}

\newpage
\clearpage

\appendix

\renewcommand\thetable{\thesection} 
\makeatletter

\makeatother

\setcounter{table}{0}  
\setcounter{figure}{0}

\section*{Supplementary Information}

\section{Details of data pre-processing for labeling the complications}
\label{apd:groundtruth-labels}
KDIGO classification was used to classify AKI encounters~\cite{khwaja2012kdigo}. The definition has three criteria, and if any of them were satisfied, the patient was assigned a diagnosis of AKI. The three criteria were either an increase in serum creatinine of 0.3 mg/dl within 48 hours, an increase of 1.5 times the baseline serum creatinine measurement, or urine output of less than 0.5 ml/kg/hr for 6 hours~\cite{khwaja2012kdigo}. We only assessed the first two definitions, since urine output was not available in our dataset. The patient's first record of serum creatinine was treated as the baseline for that patient. Patients with reported chronic kidney disease were excluded from the training and testing AKI subsets.

\begin{figure}[ht!]
\centering
\includegraphics[scale=0.6]{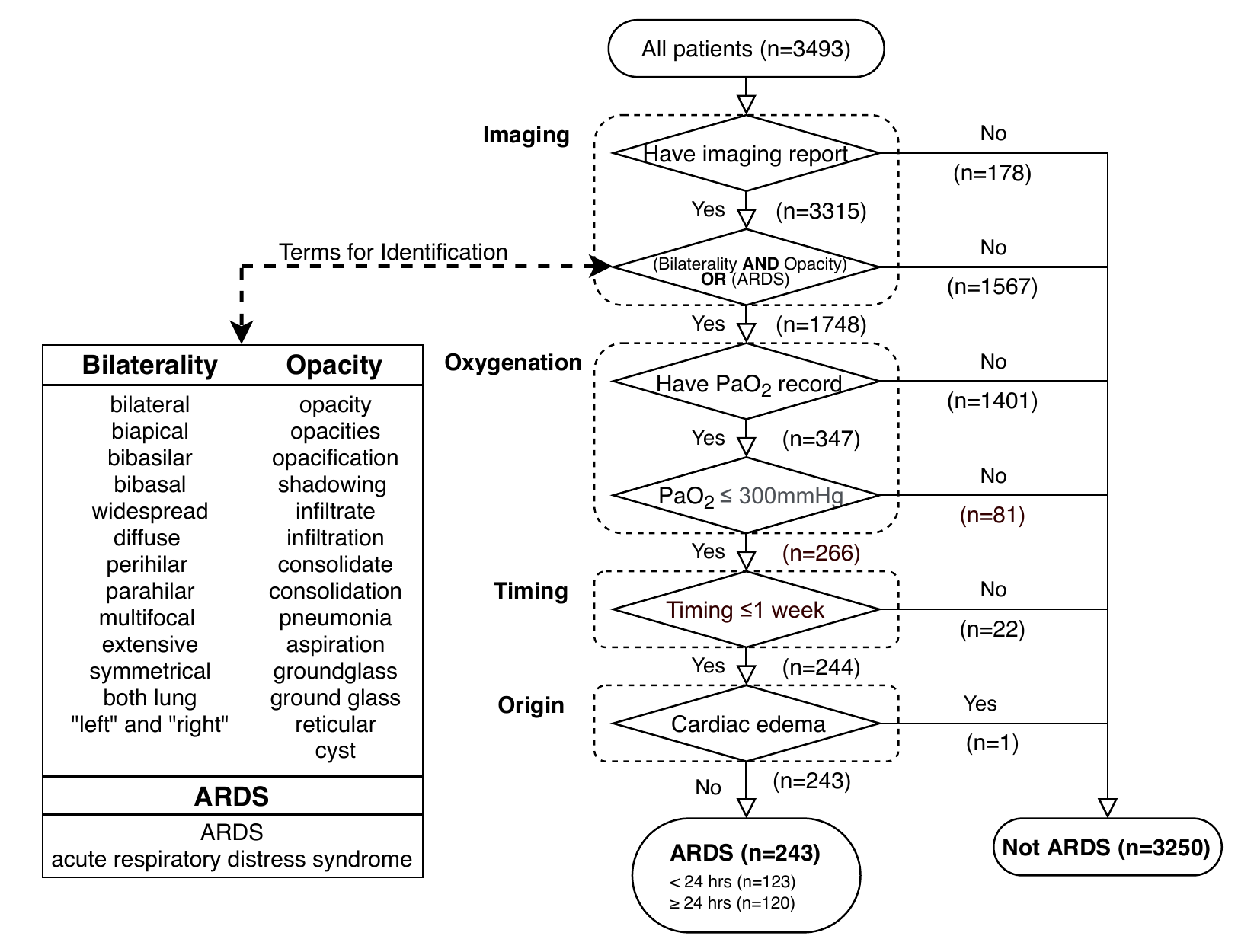}
\caption{\small The ARDS labeling process in our dataset, in accordance with the four criteria of the Berlin definition~\cite{2012ARDS_Berlin}: imaging, oxygenation, timing, and origin. The lexicon developed for identifying bilateral opacity in radiology reports is also shown within the table on the left.}
\label{fig:ARDS_labeling}
\end{figure}

The Berlin definition was employed to identify the timing and incidence of ARDS~\cite{2012ARDS_Berlin}.  The full ARDS labeling process is illustrated by the flow diagram in Figure~\ref{fig:ARDS_labeling}. Textual chest X-ray reports and CT scan reports were processed using natural language processing (NLP) techniques to identify three categorized key terms: opacity, bilaterality, and ARDS. The lexicon developed was in reference to the Herasevich~\cite{herasevich2009validation} and ASSIST~\cite{2009AzzamHelen} sniffers, which was further refined and validated based on clinical expertise. To minimize the influence of uncertainty profiles, the negation expression ``no" was searched 40 characters prior to the identification of opacity. The ARDS diagnosis was confirmed if either one of the two criteria is satisfied: (1) the ARDS term is present or (2) both terms of bilaterality and opacity are present in the report.  We identified the first radiology observation of bilateral opacity, as subsequent reports usually refer to the ones previously conducted for the identical patient instead of repeating the full interpretation and findings. Manual inspection of portions of the reports was done to validate the efficacy of the algorithm.

For the oxygenation criteria, 13,862 arterial partial pressure of oxygen ($PaO_2$) measurements acquired through arterial blood gas tests (ABG) were recorded for 358 unique patients. We have confirmed with SEHA clinicians that such test is only conducted for patients suspected of ARDS or with severe symptoms, and therefore, patients without one can be ruled out of ARDS directly. Each $PaO_2$ measurement was matched with the closet prior record of $FiO_2$ (the fraction of inspired oxygen) for the given patient to obtain the P/F ratio. For patients with missing $FiO_2$ measurements, we assumed that they were not on oxygen therapy and were assigned a value of 0.2095 (20.95\% of oxygen in air). The patients were then labeled as potentially having ARDS if their P/F ratio $\leq300$ mm Hg.

The earliest recorded time \textemdash either arrival time, admission time, or the first time the patient tested positive for COVID-19 \textemdash was utilized in lieu of the precise point of clinical insult of respiratory symptoms for the timing criteria of the Berlin definition.  To rule out pulmonary edema of other origin, patients with cardiac edema prior to the onset of ARDS were identified from the vitals and excluded. With the criteria and steps delineated herein, 243 patients were identified as having ARDS across both training sets as well as test sets.

\newpage

\section{Hyperparameter search}
\label{apd:parameters}
Our system performs random search for the hyperparameters of the machine learning models and then evaluates their performance on the validation sets. The searched hyperarameters for each of the models are shown in Table~\ref{tab:param}.

\begin{table*}[ht!]
\centering

\caption{Hyperparameter values considered during the random hyperparameter search. Ranges are indicated with a `-'.}
\begin{tabular}{ccc}
\toprule
Model &  Hyperparameters & Values \\
\midrule
&   Regularization parameter C &	[0.01, 0.10, 0.1, 10, 25, 50, 100]  \\

\textbf{Logistic Regression} &   Max iterations &	[50, 100, 200] \\

\midrule
&   Leaf size &	[1-50] \\
            
\textbf{K-Nearest Neighbors}  & Power parameter  &  [1,2]\\

&   N neighbors &	[1 - 30]  \\

\midrule
& Regularization parameter C &	[0.01, 0.10, 0.1, 10, 25, 50, 100] \\

\textbf{Support Vector Machine} & Gamma & [1e-2, 1e-3, 1e-4, 1e-5]  \\

\midrule
& L2 penalty parameter alpha & [0.005,0.002, 0.01,0.2, 0.03, 0.05] \\
&   Activation function &	[Tanh, Relu] \\
\textbf{Multi-layer Perceptron}&   Learning rate &	[Constant, Adaptive] \\

&   Weight optimization solver & [Sgd, Adam] \\

& Hidden layer sizes& [(50,50,50), (50,100,50), (100,)]\\
\midrule

&   Number of leaves &	[10-40] \\
&   Learning rate &	[0.005, 0.002, 0.01,0.2, 0.03, 0.1] \\

\textbf{Light Gradient Boosting Model}  &   Max depth &	[1-10]\\
&   N estimators &	[200-500]\\
                        
\bottomrule
\end{tabular}
\label{tab:param}
\end{table*}

\clearpage
  
\section{Model comparison}
After preprocessing the data, we compared the performance of 5 ensembles based on 5 types of base learners on the validation sets:
Logistic Regression (LR), K-Nearest Neighbors (KNN), Support Vector Machine (SVM)  Multi-layer Perceptron (MLP), and Light Gradient Boosting Model (LGBM). The models were compared using the AUROC and AUPRC and the results are shown in Table~\ref{apd:perf}. We selected the ensemble that achieved the highest AUROC on the validation set.

\begin{table*}[ht!]
\caption{\small Performance comparison for the different ensembles on the validation sets. Best performance is shown in bold. 
}
\centering
\resizebox{.8\textwidth}{!}{
\begin{tabular}{ccccccc}
\toprule
\multicolumn{2}{c}{\multirow{2}{*}{}}     & \multicolumn{2}{c}{Validation Set A} & \multicolumn{2}{c}{Validation Set B} \\
\midrule
\multicolumn{1}{c}{\textbf{Complication} } &
\multicolumn{1}{l}{\textbf{Models} } & \multicolumn{1}{c}{\textbf{AUROC}} & \multicolumn{1}{c}{\textbf{AUPRC}}  &  \multicolumn{1}{c}{\textbf{AUROC}} & \multicolumn{1}{c}{\textbf{AUPRC}}\\
\midrule

\multirow{1}{*}{} & LR & 0.908 & 0.409 & 0.977& 0.557 \\
\multirow{1}{*}{} & KNN &	0.839 & 0.298  & 0.882 &0.327\\
\multirow{1}{*}{} & SVM & 0.829 & 	0.175  & 0.956&	0.243\\

\textbf{Elevated troponin} & MLP & 0.816 &  0.201 & 0.937  & 0.291\\

& LGBM &\textbf{ 0.919} & 0.430 & \textbf{0.990 } & 0.686\\

\midrule

\multirow{1}{*}{} & LR &\textbf{ 0.742} & 0.291 & 0.724 & 0.329 \\
\multirow{1}{*}{} & KNN & 0.704 &  0.220 & 0.652 & 0.163 \\
\multirow{1}{*}{} & SVM & 0.658 & 0.198 & 0.691 & 0.217 \\

\textbf{Elevated d-dimer} & MLP & 0.700 & 0.239 & 	0.669 & 0.184\\
& LGBM & 0.741 & 0.247 & \textbf{0.753} & 0.327\\

\midrule

\multirow{1}{*}{} & LR & 0.841 & 0.361 & 0.850 & 0	0.597 \\
\multirow{1}{*}{} & KNN & 0.792 &  0.236 & 0.802 & 0.483 \\
\multirow{1}{*}{} & SVM & 0.756 & 	0.237 & 0.831 & 0.268\\

\textbf{Elevated aminotransferases} & MLP & 0.754  & 0.208  & 0.842  & 0.525 \\
& LGBM & 	\textbf{0.849}  & 0.375 & \textbf{0.884}  & 0.5786 \\

\midrule
\multirow{1}{*}{} & LR & 0.874 &  0.550& 0.905 & 0.602 \\
\multirow{1}{*}{} & KNN & 0.846 & 0.445 & 0.869 & 0.574\\
\multirow{1}{*}{} & SVM & 0.823 & 0.380  & 0.905 & 0.575 \\

\textbf{Elevated interleukin-6}& MLP &  0.853  & 0.467  & 0.881 & 0.587 \\

& LGBM &\textbf{ 0.907} & 	0.649  & \textbf{0.9464} & 0.690\\

\midrule

\multirow{1}{*}{} & LR & 0.873 & 0.360  &\textbf{ 0.945 }& 	0.625\\
\multirow{1}{*}{} & KNN & 0.783 & 0.291 & 0.918 & 0.464 \\
\multirow{1}{*}{} & SVM & .825 &  0.226 & 0.911& 0.490 \\

\textbf{SBI} & MLP & 0.802 & 0.274 & 0.935   & 	0.462 \\

& LGBM & \textbf{0.904} &  	0.501  & 0.931 & 	0.417 \\

\midrule

\multirow{1}{*}{} & LR & \textbf{0.907} & 0.459  &\textbf{ 0.829 }& 0.225\\
\multirow{1}{*}{} & KNN &	0.809 &  0.240  &0.740 &0.319 \\
\multirow{1}{*}{} & SVM & 0.818 &  0.235 & 0.807 & 0.262 \\

\textbf{AKI} & MLP & 0.775 & 0.218 & 0.809  &0.283 \\
& LGBM & 0.889  & 0.449  & 0.828 &  0.225 \\

\midrule
\multirow{1}{*}{} & LR & 0.891 &  0.310  & 	0.950 & 0.488 \\
\multirow{1}{*}{} & KNN & 0.835 & 0.178 & 0.901 & 0.364 \\
\multirow{1}{*}{} & SVM & 0.825 & 0.220  & 0.922 & 0.326 \\

\textbf{ARDS}& MLP & 0.770 & 0.143  & 0.931 &  	0.415 \\

 & LGBM & \textbf{0.911} & 0.360 &\textbf{0.960} & 0.654\\

\bottomrule
\end{tabular}}
\label{apd:perf}
\end{table*}

\end{document}